\documentclass[10pt,aps,twocolumn,superscriptaddress,showpacs,prb,reprint,floatfix,colorlinks=true,allcolors=blue]{revtex4-2}

\usepackage{amsmath, amsthm, amssymb}
\usepackage{dcolumn, bm,multirow}
\usepackage{graphicx, subfigure, verbatim}
\usepackage{latexsym}
\usepackage{braket}
\usepackage{feynmp-auto}
\usepackage{dsfont}
\usepackage{newtxtext,newtxmath}
\usepackage{xcolor}
\usepackage{soul}
\usepackage{subfiles}
\usepackage{tabularx}
\usepackage{comment}
\usepackage{hyperref}
\usepackage{kotex}
\usepackage{orcidlink}
\usepackage{mathtools}

\usepackage{xcolor}

\begin{document}

\title{Emergence of the magnetic octupole Rashba-Edelstein effect from spin-orbit entanglement}

\author{Hojun Lee\,\orcidlink{0000-0002-7406-1936}}
\affiliation{Department of Physics, Pohang University of Science and Technology, Pohang 37673, Korea}
\affiliation{Center for Quantum Dynamics of Angular Momentum, Pohang University of Science and Technology, Pohang 37673, Korea}
\author{Seungyun Han\, \orcidlink{0000-0002-6988-4751}}
\thanks{Contact Author: \href{mailto:hanson@kaist.ac.kr}{hanson@kaist.ac.kr}}
\affiliation{Department of Physics, Korea Advanced Institute of Science and Technology, Daejeon 34141, Korea}

\begin{abstract}
{Magnetic multipoles have attracted growing interest as order parameters and dynamical degrees of freedom in unconventional magnets, yet it remains unclear how broadly they can emerge as active electronic degrees of freedom. Here, we show that spin-orbit-entangled multiorbital states can host magnetic octupole (MO) degrees of freedom. In Rashba systems, this gives rise to an MO Rashba texture accompanying the $J$-Rashba texture. Remarkably, spin-orbital entanglement can realize a pure-MO limit in which the spin and orbital-angular-momentum textures vanish while the MO texture remains finite. An applied electric field converts this texture into a nonequilibrium MO polarization through an MO Rashba-Edelstein effect. Around the pure-MO regime, the MO Edelstein response dominates over the conventional spin response, showing that multipolar responses need not be small corrections to dipolar spin physics. Our results establish an interface-based route for electrically generating nonequilibrium multipolar magnetic polarization in spin-orbit-coupled systems.}
\end{abstract}

\maketitle

{\it Introduction.---} Magnetic multipoles have emerged as key descriptors of unconventional antiferromagnets whose magnetic order is invisible to the net magnetization~\cite{Hayami18PRb,Kusunose20JPSJ,Yatsushiro21PRb,Hayami21PRb,Suzuki17PRb,Smejkal22PRX2,Smejkal22PRX3,Bhowal24PRX,McClarty24PRL,Sato26npjQM,Smejkal20SCIa,Hayami21PRb,Smejkal22PRX}. In noncollinear antiferromagnets such as Mn$_3$Sn, magnetic octupoles (MOs) provide a symmetry-based language for understanding anomalous Hall responses with negligible net magnetization~\cite{Suzuki17PRb,Nakatsuji15NAT}. In altermagnets (AMs), they play an even more direct role: the compensated magnetic order 
can be characterized by higher-order magnetic multipoles, which determine the symmetry and structure of the momentum-dependent spin splitting~\cite{Bhowal24PRX,McClarty24PRL,Martinelli26PRr}. Thus, magnetic multipoles are not merely a classification scheme for complex magnetic textures; they can serve as order parameters that reveal the electronic consequences of hidden antiferromagnetic order.

Beyond their role as equilibrium order parameters, magnetic multipoles have recently emerged as dynamical degrees of freedom for controlling magnetic order. In AMs, externally injected magnetic-multipole current can exert a torque through a multipolar component symmetry-matched to the magnetic order parameter~\cite{Bhowal24PRX, McClarty24PRL, Tahir23PRL,Han25PRL,Baek25PRb,Ko25arXiv}. Such a torque is distinct from conventional spin-orbit torques~\cite{Sinova15RMP,Manchon19RMP} and provides an efficient route to electrical control of the N\'eel vector~\cite{Han26Small}.

The promise of such multipole-based physics, however, raises two fundamental questions. First, is magnetic multipole physics confined to materials with special magnetic order, or can it arise in a much broader class of materials? Second, even when present, is it necessarily subordinate to conventional magnetic-dipole physics, appearing only as a small correction to spin-based responses? Answering these questions is essential for establishing magnetic multipoles as generic and potentially dominant electronic degrees of freedom, rather than specialized descriptors of complex magnetic order.
%

In this Letter, we show that neither limitation is fundamental. Our starting point is that spin-orbit-entangled states characterized by the total angular momentum $J$ can naturally encode magnetic multipoles~\cite{Kusunose08JPSJ,Jackeli09PRL,Chen10PRb,Kusunose20JPSJ,Yang26PRb}. We demonstrate this principle in Rashba systems, where spin-orbit-entangled multiorbital states carry not only conventional spin textures but also MO textures. Since such spin-orbit-entangled Rashba states are common at inversion-asymmetric surfaces and interfaces~\cite{Bychkov84JETPL,Bihlmayer22NATrp}, this provides a broadly accessible platform for multipolar Rashba physics even in the absence of preexisting multipolar magnetic order. More remarkably, we find that spin-orbital entanglement can realize a limit in which the conventional spin and orbital-angular-momentum textures vanish while the MO texture remains finite, giving rise to a pure MO Rashba state. Finally, we show that an applied electric field converts this equilibrium MO texture into a nonequilibrium MO polarization through what we call the MO Rashba-Edelstein effect~\cite{Tahir23PRL}. Around the pure MO Rashba regime, this multipolar response dominates over the conventional spin Edelstein response, demonstrating that it need not be merely a small correction to dipolar spin physics. When coupled to a material with MO order, such as an AM, the generated MO polarization can be converted into a staggered spin density and thereby exert a N\'eel torque. The MO Rashba-Edelstein effect thus provides an interface-based route for the electrical generation and injection of MO polarization, complementary to bulk MO Hall transport~\cite{Baek25PRb,Ko25arXiv}.

\begin{figure}[t!]
\includegraphics[width=245pt]{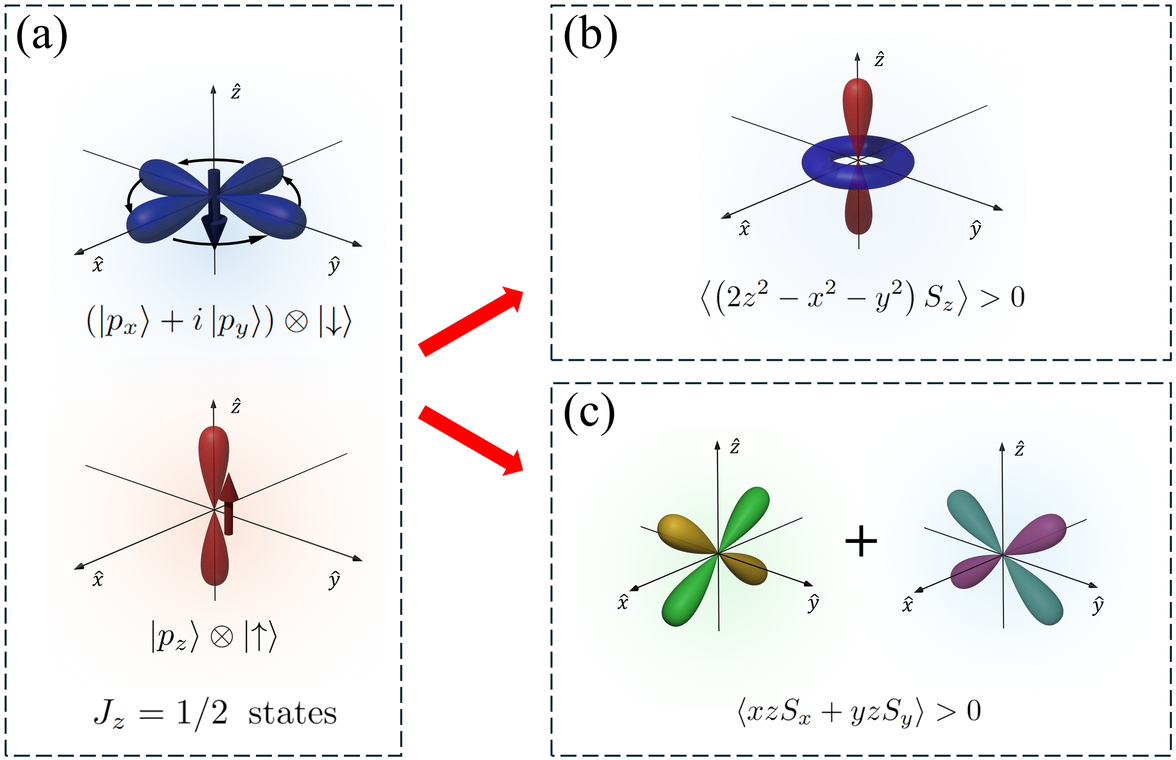}
\caption{\label{fig1} MO components encoded in a spin-orbit-entangled state. (a) The $J_z=1/2$ $p$-orbital sector comprises $(\ket{p_x}+i\ket{p_y})\otimes\ket{\downarrow}$ and $\ket{p_z}\otimes\ket{\uparrow}$. Their orbital-spin correlations yield (b) the longitudinal MO components $\langle(2z^2-x^2-y^2)\hat{S}_z\rangle>0$ and (c) the transverse MO components $\langle xz\hat{S}_x+yz\hat{S}_y\rangle>0$. Colors indicate opposite signs of the corresponding spin-density components.}
\end{figure}

{\it Magnetic multipoles encoded in $J$ states.—}
In spin-orbit-coupled systems, electronic states are commonly classified by the total angular momentum $\mathbf J=\mathbf L+\mathbf S$ and its projection, for example, $J_z= L_z + S_z$, where $S_i$ and $L_i$ are spin and orbital angular momentum operators. This classification provides a compact description of spin-orbit-entangled states, but does not exhaust their internal structure. Because spin and orbital degrees of freedom are entangled, a $J$ state contains an orbital-resolved spin distribution, or spin-orbital correlation, which can have magnetic-multipole character. Thus, a state characterized by a total-angular-momentum polarization can simultaneously carry magnetic multipole moments~\cite{Kusunose08JPSJ,Chen10PRb,Kusunose20JPSJ}.

To illustrate this point intuitively, we consider a $p$-orbital $J_z$ eigenstate [Fig.~\ref{fig1}(a)]. A $J_z=1/2$ sector of the $p$-orbital Hilbert space is spanned by $(|p_x\rangle+i|p_y\rangle)\otimes|\downarrow\rangle$ and $|p_z\rangle\otimes|\uparrow\rangle$. A spin-orbit-entangled superposition of these components has a well-defined $J_{z}$ while carrying an orbital-dependent spin distribution. The $p_x$ and $p_y$ orbitals carry predominantly down-spin character, whereas the $p_z$ orbital carries up-spin character, giving rise to a longitudinal (parallel to $S_z$) MO density of the form $(2z^2-x^2-y^2)S_z$, Fig.~\ref{fig1}(b). The spin-orbital correlations also generate transverse (orthogonal to $S_z$) MO components: the opposite spin characters of the $p_x$ and $p_z$ sectors produce an $xzS_x$ component, while those of the $p_y$ and $p_z$ sectors produce a $yzS_y$ component, Fig.~\ref{fig1}(c). Thus, the same spin-orbital entanglement that forms a $J_z$ state naturally encodes both longitudinal and transverse MO character.

This connection can be formalized by examining the operator relation between the total angular momentum and magnetic multipoles. We find that, in addition to the familiar spin and orbital angular momentum operators, higher-order magnetic-multipole operators can be constructed that commute with a given component of the total angular momentum,
\begin{align}
[J_i, M^{(n)}]=0, 
\label{eq:commutation_relation}
\end{align}
where $M^{(n)}$ is an $n$th-order Cartesian magnetic-multipole operator. For a given $J_i$, several compatible multipoles can be constructed, including the longitudinal and transverse components above. Their construction and higher-order generalization are presented in Supplemental Material~\cite{suppl_ref}. Compatibility alone does not guarantee a finite expectation value in a given manifold; this is determined by whether the projected operator $PM^{(n)}P$ remains finite.

For the MO Rashba effect, we use the vector-like Cartesian MO operator $\mathbf O$~\cite{Hayami21PRb} which satisfies Eq.~\eqref{eq:commutation_relation}, and provides a convenient basis for characterizing the Rashba texture~\footnote{This choice does not imply that the multipolar content of a $J$ state is
restricted to this vector-like channel; higher-rank magnetic-multipole
components can in general coexist}:
\begin{align}
\mathbf{O}
=
\frac{2}{\sqrt{10}}
\left[
3(\mathbf r\cdot\mathbf S)\mathbf r-r^2\mathbf S
\right].
\label{eq:D_def}
\end{align}
Here, MO denotes the broader Cartesian magnetic moment $r_i r_j S_k$, a quadrupolar spatial distribution coupled to a spin (see Supplemental Material~\cite{suppl_ref}). Each $O_i$ combines longitudinal and transverse MO channels compatible with $J_i$. It directly characterizes the MO polarization of $J$-Rashba states. Numerically, $\mathbf O$ is evaluated in its atomic-orbital representation given in Supplemental Material~\cite{suppl_ref}.

\begin{figure}[t!]
\includegraphics[width=0.45\textwidth]{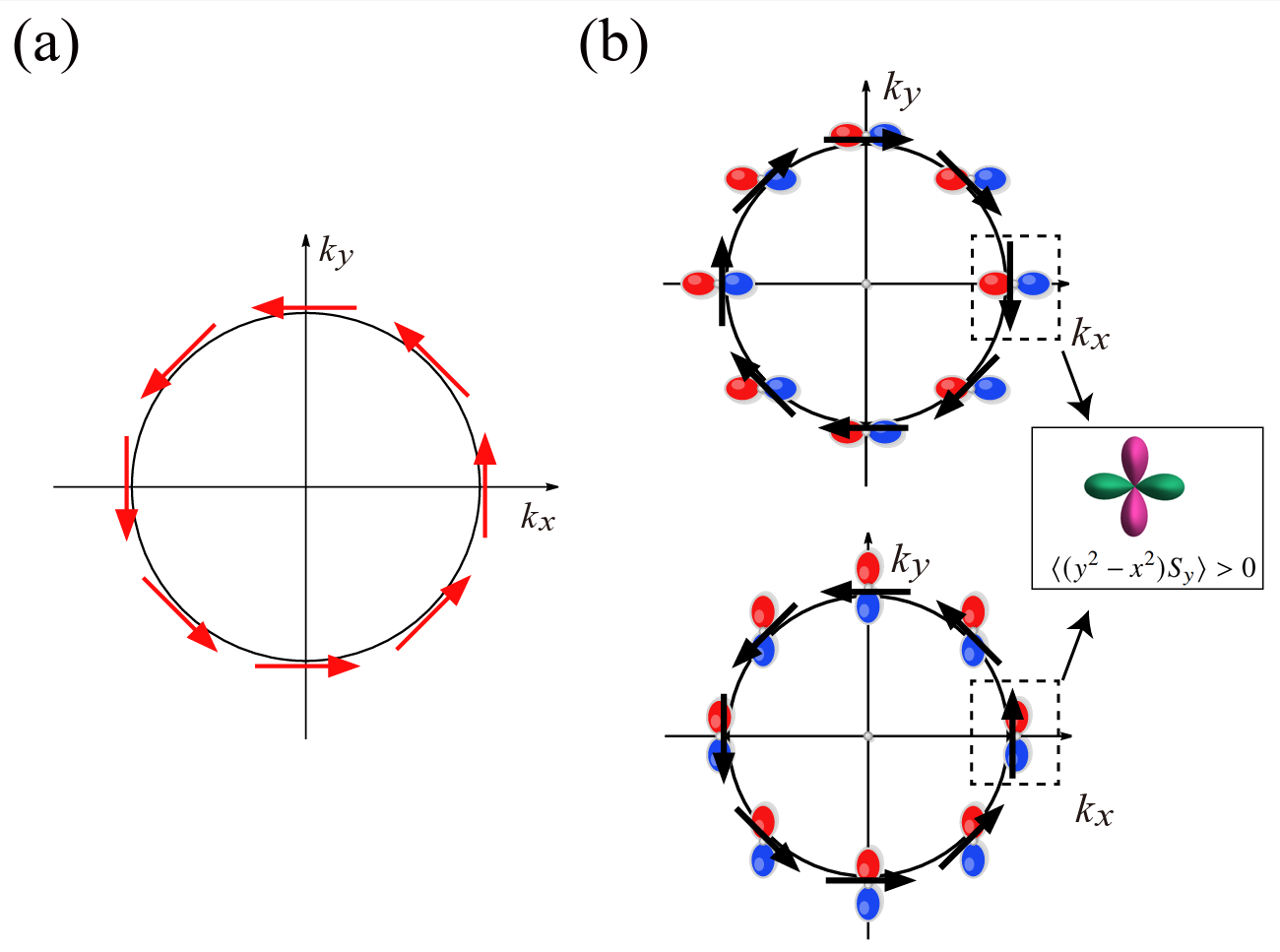}
\caption{\label{fig2}
Emergence of a pure MO Rashba texture from a $J$-Rashba state.
(a) Generic $J$-Rashba state.
(b) In the pure-MO limit, the $J$ texture is encoded in
orbital-resolved spin textures with opposite spin polarizations.
Their contributions cancel in the net spin channel but add in
the MO channel, yielding a finite MO Rashba texture with
vanishing spin texture.
Dashed boxes mark the same sign of $O_{y}$ for $k_{x}>0$; green/purple denote opposite $S_{y}$ densities.}
\end{figure}

{\it Magnetic-octupole Rashba systems.—}
We now show that a $J$-Rashba texture is generically accompanied by a MO
Rashba texture and can even enter a regime where the latter dominates over
the conventional spin texture.
In spin-orbit-entangled bands, inversion-symmetry breaking can generate a
Rashba coupling to the total angular momentum~\cite{Bychkov84JETPL,Kim13PRb,Bihlmayer22NATrp,Bahari26PRb} [Fig.~\ref{fig2}(a)], 
\begin{align}
H_R=\alpha_R(\mathbf k \times \hat{\mathbf z})\cdot \mathbf J,
\end{align}
where $\alpha_R$ is the Rashba constant. The resulting Rashba eigenstates carry a momentum-dependent $J$ polarization. Since a $J$-polarized state can simultaneously possess finite spin and MO moments, $\langle\mathbf S\rangle$ and $\langle\mathbf O\rangle$, the $J$-Rashba texture can manifest as both spin and MO textures.

Importantly, the spin and MO textures need not follow the same behavior in crystalline systems. In the rotationally symmetric atomic limit, a $J$ state can carry both conventional angular-momentum and MO characters. In a crystal, however, crystal-field splitting reduces the orbital manifold and modifies the internal spin-orbital structure of the Rashba states. 
This projection can naturally lead to a regime in which the in-plane spin and orbital-angular-momentum textures vanish, $\langle\mathbf S_{\parallel}\rangle=\langle\mathbf L_{\parallel}\rangle=0$, while the MO texture remains finite, $\langle\mathbf O_{\parallel}\rangle\neq0$ [Fig.~\ref{fig2}(b)],
\begin{align}
H_R=\alpha_R^{\prime}(\mathbf k \times \hat{\mathbf z})\cdot \mathbf O.
\end{align}
The resulting state represents a pure MO Rashba state, as we demonstrate explicitly below.

{\it Magnetic-octupole Rashba states in a $p$-orbital model.---}
We now demonstrate that MO Rashba states emerge naturally in a minimal model of conventional Rashba systems. We consider $p_x$, $p_y$, and $p_z$ orbitals on a two-dimensional triangular lattice,
\begin{align}
H
&=
H_{\rm NN}
+\left[ V_{\rm ISB}
\sum_{\langle ij\rangle,s}
\sum_{\mu=x,y}
\hat{\delta}_{ij,\mu}
\left(
c^{\dagger}_{i\mu s}c_{j z s}
-
c^{\dagger}_{i z s}c_{j\mu s}
\right) + \text{h.c.} \right]
\nonumber\\
&\quad
+\Delta_{\rm CF}\sum_{i,s}
c^{\dagger}_{i z s}c_{i z s}
+\frac{\lambda_{\rm soc}}{\hbar^{2}}
\sum_i
\sum_{\mu,\nu}
\sum_{s,s'}
c^{\dagger}_{i\mu s}
\left(
\mathbf L_{\mu\nu}\cdot \mathbf S_{ss'}
\right)
c_{i\nu s'},
\label{eq:p_orbital_real_space}
\end{align}
where $(i,j)$, $(s,s')$, and $(\mu,\nu)$ are site, spin, and orbital
indices, $\langle i,j\rangle$ denotes nearest neighbors, and $\hat{\delta}_{ij,\mu}$ is the $\mu$ component of the unit bond vector from site $i$ to $j$. The model contains only standard orbital/spin Rashba ingredients: nearest-neighbor hopping ($H_{\rm NN}$), interface-induced crystal-field splitting ($\Delta_{\rm CF}$), ISB interorbital hopping ($V_{\rm ISB}$), and atomic spin-orbit coupling (SOC)~\cite{Go17SR}, with no preexisting magnetic-multipole order or additional multipolar interaction (see Supplemental Material~\cite{suppl_ref}).

\begin{figure*}[t!]
\includegraphics[width=500pt]{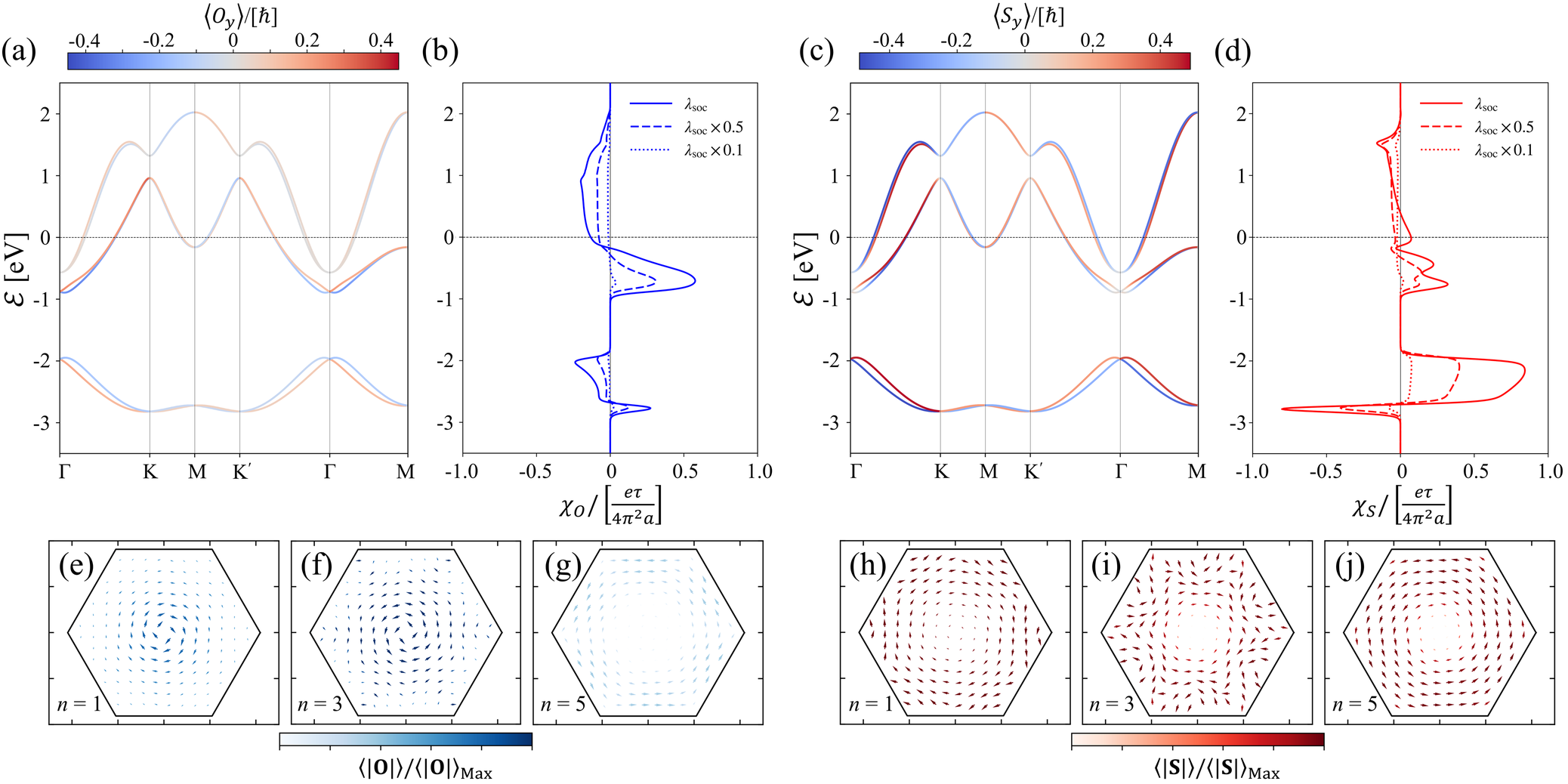}
\caption{\label{fig3}Magnetic-octupole and spin Rashba textures and Rashba-Edelstein responses in the $p$-orbital model. (a),(c) Bands colored by $\braket{O_{y}}$ and $\braket{S_{y}}$, respectively. (b),(d) MO and spin Rashba-Edelstein susceptibilities, $\chi_{O}=\delta \braket{O_{y}}/E_{x}$ and $\chi_{S}=\delta \braket{S_{y}}/E_{x}$ versus energy $\mathcal{E}$. Solid, dashed, and dotted lines correspond to $\lambda_{\text{soc}}$, $0.5 \lambda_{\text{soc}}$, and $0.1 \lambda_{\text{soc}}$. Here, $\tau$ is the relaxation time and $a$ the lattice constant. (e)-(g), (h)-(j) Momentum-space in-plane MO and spin textures. $n$ denotes the band index; arrows show $\braket{\mathbf{O}_{\parallel}}$ and $\braket{\mathbf{S}_{\parallel}}$, and color intensities their magnitudes normalized by the common maximum.}
\end{figure*}

At $\Gamma$, the crystal field separates $p_z$ from the $p_x$-$p_y$ manifold, while SOC further organizes the bands into $|J_{z}|=3/2$ (bands 5 and 6), $|J_{z}|=1/2$ (bands 3 and 4), and predominantly $p_z$ (bands 1 and 2) sectors [Fig.~\ref{fig3}(a)].
Expanding around the $\Gamma$ point yields
\begin{align}
H_{|J_{z}|=1/2}
\propto 
\alpha_R^{|J_{z}|=1/2}
(\mathbf{k}\times\hat{\mathbf z})
\cdot
\boldsymbol{\sigma},
\end{align}
for the $|J_{z}|=1/2$ sector, and
\begin{align}
H_{p_z}
\propto
\alpha_R^{p_z}
(\mathbf{k}\times\hat{\mathbf z})
\cdot
\boldsymbol{\rho},
\end{align}
for the predominantly $p_z$ sector. Here, $\bm{\sigma}$ and $\bm{\rho}$ are Pauli matrices representing the low-energy pseudospins of the $|J_{z}|=1/2$ and predominantly $p_{z}$ sectors, respectively. Although these Rashba bands usually discussed from the viewpoint of spin textures, they exhibit qualitatively different properties when viewed in terms of magnetic multipoles.

We first consider the isolated $|J_{z}|=1/2$ sector with pseudospin basis
$\ket{u}=(\ket{p_x}+i\ket{p_y})\otimes\ket{\downarrow}/\sqrt{2}$ and
$\ket{d}=(\ket{p_x}-i\ket{p_y})\otimes\ket{\uparrow}/\sqrt{2}$.
Projecting physical operators onto this subspace yields the striking result
\begin{align}
P_{1/2} L_i P_{1/2} = P_{1/2} S_i P_{1/2} = 0,
\qquad
P_{1/2} O_i P_{1/2} \propto \sigma_i,
\end{align}
for $i=x,y$. Importantly, the absence of the conventional spin and orbital-angular-momentum textures follows at the operator level within the projected doublet, rather than from an accidental cancellation on a particular Fermi contour. Thus, the Rashba pseudospin carries no conventional magnetic-dipole degree of freedom while retaining a finite MO component. We refer to this regime as a \emph{pure MO Rashba state}. The microscopic origin of this seemingly counterintuitive result becomes transparent from the eigenstates of $\sigma_i$. Consider, for instance, the $\sigma_y=+1$ state, $(\ket{u}+i\ket{d})/\sqrt{2}\propto
\ket{p_x} \otimes (\ket{\downarrow}+i\ket{\uparrow})
+i \ket{p_y} \otimes (\ket{\downarrow}-i\ket{\uparrow})$, which can be expressed as 
$\ket{p_x} \otimes \ket{S_y=-}
+i \ket{p_y} \otimes \ket{S_y=+}$. The $p_x$ and $p_y$ orbitals therefore carry opposite $S_y$ polarizations, Fig.~\ref{fig2}(b). Their contributions cancel in the net spin density, $\langle S_y\rangle=0$, but remain finite when resolved by the orbital character. In real space, this corresponds to opposite $S_y$ densities distributed along the $x$ and $y$ directions, producing a finite MO density with $(y^2-x^2)S_y$ symmetry, Fig.~\ref{fig2}(b). The opposite eigenstate of $\sigma_y$ reverses this MO density while still carrying zero net spin. Thus, the cancellation of the orbital-resolved spin polarizations removes the net spin texture while preserving their spin-orbital correlation as a finite MO. Consequently, the Rashba winding survives in the MO channel, as illustrated in Fig.~\ref{fig2}. 

The predominantly $p_z$ sector exhibits a qualitatively different behavior.
Its pseudospin basis is simply
$(\ket{p_z}\otimes\ket{\uparrow},
\ket{p_z}\otimes\ket{\downarrow})$,
for which
\begin{align}
P_{z} S_i P_{z} \propto\rho_i,
\qquad
P_{z} O_i P_{z} \propto\rho_i,
\end{align}
where $P_{z}$ is the projection operator onto the predominantly $p_{z}$ sector. Thus, the Rashba state simultaneously carries both spin and MO textures.
The MO component originates from the anisotropic spatial profile of the spin-polarized $p_z$ orbital.
For example, a $p_z$ orbital with spin polarized along the $y$ direction simultaneously carries the MO component $(2z^2-x^2-y^2)S_y$.
Hence, the conventional spin-Rashba state should be viewed not as a purely spin texture, but as the coexistence of spin and MO textures (see Supplemental Material~\cite{suppl_ref}).

We next perform tight-binding calculations to examine how the MO Rashba texture manifests in the full electronic structure. Figure~\ref{fig3} shows the band-resolved $\mathbf{O}$ and $\mathbf{S}$ Rashba textures. A finite $\mathbf{O}$ texture appears broadly across the Rashba-split bands, consistent with the MO character encoded in the underlying $J$ states. Near the $\Gamma$ point, the three orbital manifolds exhibit distinct behaviors. In bands 3 and 4, the $\mathbf{O}$ texture is pronounced while the $\mathbf{S}$ texture is suppressed, providing a direct manifestation of the pure MO Rashba state discussed above. In contrast, both textures are weak in bands 5 and 6, whereas bands 1 and 2 exhibit both $\mathbf{O}$ and $\mathbf{S}$ textures, as expected from their simultaneous spin and MO character. These results demonstrate that the multipolar Rashba texture persists beyond the effective low-energy description and is broadly distributed throughout the Rashba electronic structure.

Importantly, the MO Rashba texture is not merely a multipolar reinterpretation of the conventional Rashba state, but gives rise to a distinct nonequilibrium response. An applied electric field converts the momentum-space MO texture into a finite nonequilibrium MO polarization through the Rashba-Edelstein effect~\cite{Edelstein90SSC,Manchon08PRb,Yoda18NL,Tahir23PRL}. For an electric field $E_x$, the induced expectation value of an operator $D$ is evaluated within the constant-relaxation-time intraband approximation as
\begin{align}
\delta\langle D_y\rangle
=\chi_D E_x,
\quad
\chi_{D}=
-e\tau
\sum_{n\mathbf{k}}
\left(-\frac{\partial f}{\partial\epsilon}\right)
\langle D\rangle_{n\mathbf{k}}
v_{n\mathbf{k},x},
\label{eq:edelstein}
\end{align}
where $\tau$ is the relaxation time, $v_{n\mathbf{k},x}$ is the group velocity, and $D$ denotes either $O_y$ or $S_y$. Figures~\ref{fig3}(b) and \ref{fig3}(d) show the resulting $\chi_{O}$ and $\chi_{S}$ as functions of the Fermi energy. A finite MO accumulation appears over a broad energy range, directly reflecting the widespread $O_y$ Rashba texture in the band structure. More importantly, the relative strengths of the MO and spin Edelstein responses closely follow their underlying Rashba textures. Around bands 3 and 4, the MO accumulation dominates over the spin accumulation. Both responses become weak around bands 5 and 6, whereas both densities are present in the predominantly $p_z$ bands. Thus, the MO character encoded in the equilibrium Rashba texture is directly converted into a nonequilibrium MO density by an applied electric field.

Material calculations for BiAg$_2$, BiTeI, and
LaAlO$_3$/SrTiO$_3$ reveal MO Edelstein responses comparable to their spin
counterparts (see Supplemental Material~\cite{suppl_ref}).
Since these are prototypical Rashba platforms with sizable spin responses,
this result suggests that MO responses can be similarly strong in realistic
systems, with MO-dominant regimes potentially accessible through materials
exploration.

{\it Application: N\'eel magnetic-octupole torque on $d$-wave AMs.—}
The nonequilibrium MO density generated by the MO Rashba-Edelstein effect provides a route to magnetic dynamics distinct from conventional spin accumulation.
A particularly natural setting is a $d$-wave AM, whose N\'eel order itself carries MO character.
Previous studies have shown that a MO current can exert a torque on such multipolar magnetic order~\cite{Han25PRL,Han26Small}.
Here, we show that the MO Rashba-Edelstein effect provides a direct electrical route to the same multipolar degree of freedom within the AM itself. Once inversion symmetry is broken in a $d$-wave AM, an MO Rashba state can emerge and an electric field naturally generates a nonequilibrium MO density accompanied by a sublattice-staggered spin polarization.
We explicitly demonstrate this mechanism using a minimal inversion-asymmetric $d$-wave AM model in the End Matter.

The resulting staggered spin density exerts a field-like torque on the N\'eel order, which we call the N\'eel MO torque (NMOT):
\begin{align}
\mathbf{T}_{\text{MO}}
=T_{\text{MO,FL}}
\mathbf{\hat{n}}\times(\mathbf{\hat{z}}\times \mathbf{E}), 
\label{eq:Neel_MOT}
\end{align}
where $T_{\text{MO,FL}}$ is the field-like torque coefficient proportional to the nonequilibrium MO density [$\delta \mathbf{O} = \chi_O(\hat{\mathbf{z}}\times \mathbf{E})$]~\cite{Han25PRL}. The NMOT is reminiscent of the N\'eel spin-orbit torque in CuMnAs~\cite{ZeleznyPRL2014,Zelezny17PRb,Salemi19NATc}, where a current-induced staggered spin polarization enables electrical reorientation of the N\'eel vector.
Here, however, the staggered response arises from the MO Rashba-Edelstein effect together with the sublattice-dependent orbital character of the $d$-wave AM. Thus, inversion-asymmetric AMs, such as GdAlSi~\cite{nag2024}, naturally provide a platform for generating a NMOT without requiring an external source of MO polarization. In addition, NMOT may be particularly relevant to surface AMs~\cite{lange2026,hu26PRL}, which have recently attracted growing interest and naturally host inversion-asymmetric environments due to surface-induced symmetry breaking.

{\it Discussion and outlook.---} Our work establishes a connection between total-angular-momentum ($J$) physics and magnetic-multipole physics, thereby extending the scope of magnetic-multipole phenomena beyond systems conventionally described by multipolar order.
This connection provides a new perspective on several known spin-orbit-entangled states. Indeed, the established surface states of Bi$_2$Se$_3$ and Bi$_2$Te$_3$~\cite{Zhang13PRL,Zhu13PRL,Xie14NATc} already encode a helical MO Rashba texture, $\propto (\mathbf{k} \times\mathbf O)\cdot \hat{\mathbf z}$, as shown analytically in the Supplemental Material~\cite{suppl_ref}. This demonstrates that the MO physics discussed here naturally occurs in prototypical spin-orbit-coupled surface states. This connection further suggests an interplay between topology and magnetic-multipole physics.
For example, heterostructures combining Bi$_2$Se$_3$ with multipolar magnets, such as AMs or Mn$_3$Sn, may provide a platform for exploring electrically driven multipolar magnetic dynamics induced by topological surface states. 

From a complementary perspective, multipolar pseudospins with suppressed conventional spin components have been identified in spin-orbit-entangled Kramers doublets of a strongly correlated $d^1$ Mott insulator~\cite{Jackeli09PRL}.
Our results show that closely related multipolar $J$ physics can emerge in the much more common setting of inversion-asymmetric Rashba systems.

The predicted MO Rashba physics also provides experimentally accessible signatures.
The MO Rashba texture can be probed by spin- and angle-resolved photoemission spectroscopy, where polarization-dependent photoemission can resolve the orbital-dependent spin textures constituting the MO polarization.
Indeed, existing observations of orbital-polarized spin textures in topological surface states may already contain signatures of such a texture~\cite{Xie14NATc,Bentmann21PRb}.
For the nonequilibrium response, the NMOT predicted in inversion-asymmetric AMs may enable reorientation of the N\'eel vector, analogous to the N\'eel spin-orbit torque in CuMnAs~\cite{ZeleznyPRL2014,Zelezny17PRb}, which could be detected through the corresponding anisotropic magnetoresistance~\cite{Wadley16SCI}. 

{\it Note added.---} During the final stages of preparing this manuscript, we became aware of a recent preprint~\cite{Jo26arXiv} on how altermagnetic nonrelativistic spin splitting reshapes spin Rashba physics in the $d$-orbital square lattice. That work focuses on nonrelativistic modification of the spin-Rashba response, whereas the present work addresses the magnetic-multipole Rashba degree of freedom and its Edelstein response.

{\it Acknowledgments.---} We thank Hyun-Woo Lee, Kyoung-Whan Kim, Suik Cheon, Insu Baek, and Jeonghun Sohn for fruitful discussions. H. Lee and S. Han were financially supported by the National Research Foundation of Korea (NRF) grant funded by the Korean government (MSIT) (No. RS-2024-00356270, RS-2024-00410027). S. Han were financially supported by the InnoCORE program of the Ministry of Science and ICT(N10260140).

\begin{verbatim}
\end{verbatim}

\nocite{*}

\bibliography{Paper}

\newpage
\onecolumngrid
\begin{center}
  \textbf{End Matter}
\end{center}

\twocolumngrid
\clearpage

{\it Magnetic-octupole Rashba-Edelstein effect in $d$-wave altermagnets.---} We demonstrate that the MO Rashba-Edelstein effect can generate a net torque in a $d$-wave AM.
Specifically, inversion-symmetry breaking (ISB) generates a MO Rashba-Edelstein response, which, through the sublattice-dependent orbital character of the AM, produces a staggered spin density and a net field-like torque on the N\'eel vector. To demonstrate this, we consider a minimal square-lattice model consisting of two magnetic sublattices, A and B, and three $t_{2g}$ orbitals, $d_{xy}$, $d_{yz}$, and $d_{zx}$~\cite{Jo25PRL}. In the basis
$\{A, B\} \otimes \{ d_{xy}, d_{yz}, d_{zx} \} \otimes \{\ket{S_{z} = \uparrow}, \ket{S_{z} = \downarrow}\}$
the Hamiltonian is written in the block form
\begin{align}
H(\mathbf k)= \left(
\begin{array}{cc}
    H_A(\mathbf k) & T(\mathbf k) \\
    T^{\dagger}(\mathbf k) & H_B(\mathbf k)
\end{array} \right) .
\label{eq:AM_H}
\end{align}
The intersublattice hopping is diagonal in the orbital and spin indices,
\begin{align}
T(\mathbf k)=
{\rm diag}
\left(
h_{xy},h_{xy},
h_{yz},h_{yz},
h_{zx},h_{zx}
\right),
\end{align}
where $h_{xy}=2t_{\pi}(c_x+c_y)$, $h_{yz}=2(t_{\delta}c_x+t_{\pi}c_y)$, and $h_{zx}=2(t_{\pi}c_x+t_{\delta}c_y)$ with $c_i=\cos(k_i d)$, $d=a/\sqrt{2}$, and $a$ is the lattice constant.
%
\begin{figure}[t!]
\includegraphics[width=245pt]{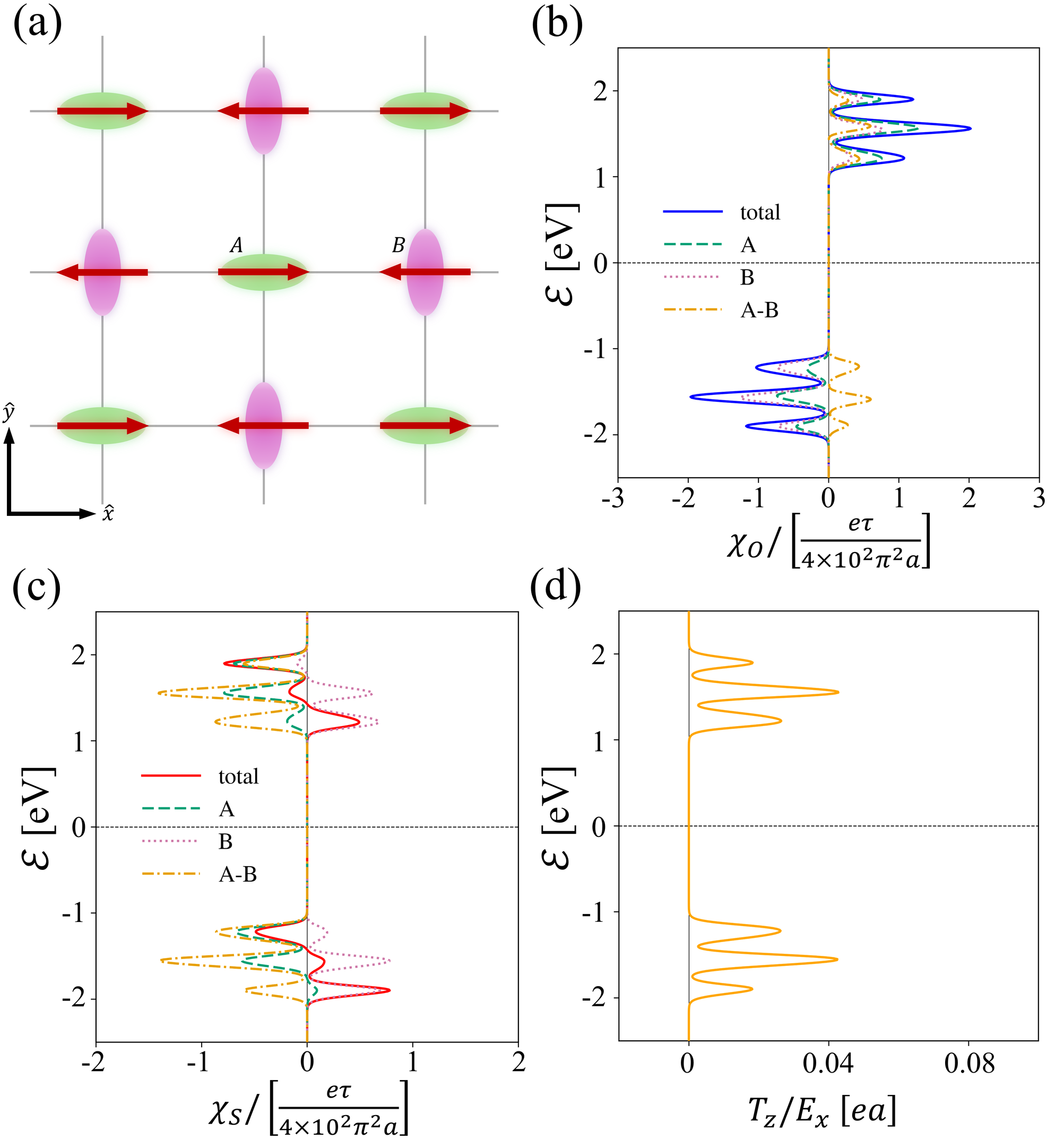}
\caption{\label{fig4} Magnetic-octupole Rashba-Edelstein effect in a $d$-wave AM.
(a) Schematic of the two-sublattice model with sublattice-dependent orbital character.
(b), (c) Total and sublattice-resolved MO and spin Rashba-Edelstein susceptibilities, respectively.
The $A-B$ component denotes the staggered response.
(d) Current-induced torque response. $T_{z}$ denotes the $z$ component of the current-induced torque. For (d), we use $\Gamma = \hbar / \tau = 25$ meV.}
\end{figure}
%
The sublattice Hamiltonian blocks ($H_{A}$, $H_{B}$) contain the $s-d$ exchange, crystal-field splitting, atomic SOC, and ISB orbital hopping. We parameterize the SOC and crystal-field strengths by $\lambda_{\rm soc}$ and $\Delta_{\rm CF}$, respectively. The staggered exchange field is described by $J_{sd}$ with the in-plane N\'eel vector $\mathbf N=(\cos\phi,\sin\phi,0)$, where $\phi$ is the azimuthal angle. Importantly, the crystal-field term also changes sign between the two sublattices, producing the sublattice-dependent orbital character characteristic of the $d$-wave AM (see Supplemental Material~\cite{suppl_ref} for details). Broken inversion symmetry is introduced through the intra-sublattice orbital hopping characterized by
\begin{align}
g_x&=2\lambda_{\rm ISB}s_xc_y,\qquad
g_y=2\lambda_{\rm ISB}c_xs_y,
\end{align}
where $s_i=\sin(k_i d)$.
Together with atomic SOC, this ISB hopping generates spin and MO Rashba textures. The intra-sublattice ISB hopping phenomenologically captures local inversion asymmetry around each magnetic sublattice; the same mechanism persists for intersublattice ISB hopping~\cite{suppl_ref}. The explicit form of the sublattice Hamiltonians $H_A$ and $H_B$, together with further details of the model, is given in the Supplemental Material~\cite{suppl_ref}. In the following, we focus on the representative configuration $\mathbf{N}\parallel\hat{\mathbf{x}}$ and $\mathbf{E}\parallel\hat{\mathbf{x}}$. 

We calculate the nonequilibrium spin and MO densities induced by an electric field $E_x$.
Figures~\ref{fig4}(b) and \ref{fig4}(c) show the sublattice-resolved Rashba-Edelstein susceptibilities together with their total and staggered components.
The latter are obtained by taking the sum and difference of the $A$- and $B$-sublattice responses, $\chi_{D,\mathrm{tot}}=\chi_{D,A}+\chi_{D,B}$ and $\chi_{D,A-B}=\chi_{D,A}-\chi_{D,B}$, respectively, where $\chi_{D,i}=\delta\langle D_{y,i}\rangle/E_x$ with $D=O,S$.

Both the MO and spin Rashba-Edelstein responses are finite, reflecting the
coexistence of MO and spin Rashba textures in this system~\cite{suppl_ref}.
Importantly, in addition to the net spin response, a sizable staggered spin
density $\chi_{S,A-B}$ emerges [dash-dotted orange line in Fig.~\ref{fig4}(c)].
We find that this staggered response results from the combined effect of the
nonequilibrium MO polarization and the sublattice-dependent orbital character
of the $d$-wave AM.

To substantiate this interpretation, we reverse the staggered crystal field,
which interchanges the dominant orbital characters of the two sublattices.
While the electric-field-induced MO polarization remains unchanged, the
staggered spin response reverses its sign,
\begin{align}
\Delta_{\mathrm{CF}}\rightarrow-\Delta_{\mathrm{CF}}:\qquad
\chi_{O,\mathrm{tot}}
&\rightarrow\chi_{O,\mathrm{tot}},&
\chi_{S,A-B}
&\rightarrow-\chi_{S,A-B}.
\end{align}
Moreover, in the limit $\Delta_{\mathrm{CF}}\rightarrow0$, the MO response
remains finite whereas the staggered spin response vanishes.
These results indicate that the MO Rashba-Edelstein polarization contributes
to the staggered spin response through the sublattice-dependent orbital
structure of the $d$-wave AM (see Supplemental Material~\cite{suppl_ref}).

This behavior can be understood from the orbital-resolved picture of the MO
Rashba-Edelstein effect.
The nonequilibrium MO polarization generated by the electric field encodes
an orbital-dependent spin polarization, as illustrated in Fig.~\ref{fig2}(b).
While this picture was introduced using the $p_x$ and $p_y$ orbitals, the same
mechanism applies here to the $d_{yz}$ and $d_{zx}$ orbitals.
In the $d$-wave AM, the staggered crystal field locks opposite
orbital characters to the two magnetic sublattices.
Consequently, the orbital-dependent spin polarization associated with the MO
response acquires opposite signs on the two sublattices, thereby contributing
to a staggered spin density.

The resulting staggered spin density exerts a field-like torque on the
N\'eel order through the $s$-$d$ exchange.
The contribution associated with the nonequilibrium MO polarization can be
parameterized as~\cite{Han25PRL}
\begin{align}
\mathbf{T}_{\mathrm{MO}}
=
T_{\mathrm{MO,FL}}\,
\hat{\mathbf n}\times
(\hat{\mathbf z}\times\mathbf E),
\label{eq:Neel_MOT}
\end{align}
where $T_{\mathrm{MO,FL}}$ denotes the coefficient of the field-like torque
associated with the nonequilibrium MO polarization
$\delta\mathbf O=\chi_O(\hat{\mathbf z}\times\mathbf E)$.
For $\mathbf E\parallel\hat{\mathbf x}$ and
$\mathbf N\parallel\hat{\mathbf x}$, the induced $\delta O_y$ therefore
contributes to a net field-like torque along $\hat{\mathbf z}$. Figure~\ref{fig4}(d) shows the corresponding current-induced torque response,
which becomes sizable in the same energy region where the MO
Rashba-Edelstein response is appreciable. This mechanism is reminiscent of the N\'eel spin-orbit torque in CuMnAs,
where a staggered current-induced spin polarization exerts a torque on the
N\'eel order~\cite{ZeleznyPRL2014,Zelezny17PRb}.
Here, the MO Rashba-Edelstein polarization, together with the
sublattice-dependent orbital character of the $d$-wave AM,
provides an additional route to generating the staggered spin polarization
and the resulting field-like torque.
We refer to this MO-associated torque contribution as the N\'eel magnetic-octupole torque (NMOT).

\onecolumngrid

\newpage

\appendix

\begin{center}
  \large{\textbf{Supplemental Material for ``Emergence of the magnetic octupole Rashba-Edelstein effect from spin-orbit entanglement"}}
\end{center}

\tableofcontents
\appendix


\section{Magnetic multipole operators}\label{sec:Atomic_MO_operators}

In this section, we clarify the convention for magnetic multipole operators used throughout the present work. In the Cartesian representation, an $n$th-order local magnetic multipole can be represented by a tensor of the form
\begin{align}
    \mathcal{M}^{(n)} = r_{i_{1}} \cdots r_{i_{n-1}} \mu_{i_{n}},
\end{align}
where $\mathbf{r}$ denotes the position measured from the multipole center and $\bm{\mu}$ is the local magnetic moment. In the present work, we focus on the spin moment and take $\bm{\mu} \propto \mathbf{S}$, with the overall prefactor absorbed into the normalization of the multipole operator. That is, we employ the magnetic multipole operator of the form
\begin{align}
    M^{(n)} = r_{i_{1}} \cdots r_{i_{n-1}} S_{i_{n}}.
\end{align}

The superscript $n$ in $M^{(n)}$ denotes the Cartesian multipole order. We emphasize that a Cartesian tensor of order $n$ is generally reducible under spatial rotations to irreducible spherical tensors of rank $\leq n$~\cite{Kusunose20JPSJ}. Throughout this work, we use the term ``magnetic multipole" primarily in this Cartesian-tensor sense.

\subsection{Magnetic octupoles}

As the central example of the present work, we consider the third-order Cartesian magnetic multipole (magnetic octupole)
\begin{align}\label{eq:MO_operator}
    M^{(3)} = r_{i}r_{j}S_{k}.
\end{align}
Since the position operators commute, the magnetic octupole (MO) is symmetric on exchange of $i$ and $j$. This symmetry reduces the number of independent parameters from 27 to 18. Under this restriction, the decomposition of the direct product of three vector representations is reduced from
\begin{align}
    3 \otimes 3 \otimes 3 = 1 \oplus 3 \oplus 3 \oplus 3 \oplus 5 \oplus 5 \oplus 7
\end{align}
to
\begin{align}
    (3 \otimes 3)_{\text{sym}} \otimes 3 = (1 \oplus 5) \otimes 3 = (3 \oplus 3) \oplus 5 \oplus 7.
\end{align}
If we restrict the spatial quadrupole to its traceless part by introducing
\begin{align}\label{eq:quadrupole_sym}
    Q_{ij} = r_{i}r_{j} - \frac{1}{3}\delta_{ij}r^{2},
\end{align}
then the decomposition is reduced to
\begin{align}
    3 \oplus 5 \oplus 7.
\end{align}
These three irreducible sectors correspond to the rank-one anisotropic magnetic dipole (AMD), the rank-two magnetic toroidal quadrupole (MTQ), and the rank-three reduced magnetic octupole (RMO), respectively~\cite{Urru22AOP,Sato26npjQM}.

In the present work, we focus on the vector-like rank-one Cartesian MO. Contracting the quadrupole tensor in Eq.~\eqref{eq:quadrupole_sym} with the spin gives
\begin{align}
    D_{i} = 3Q_{ij}S_{j} = 3(\mathbf{r} \cdot \mathbf{S})r_{i} - r^{2} S_{i},
\end{align}
which is conventionally referred to as the AMD~\cite{Kusunose20JPSJ,Hayami21PRb}. We define the normalized vector-like MO operator used in the main text as
\begin{align}\label{eq:D_def}
    \mathbf{O} = \frac{2}{\sqrt{10}} \left[ 3(\mathbf r\cdot\mathbf S)\mathbf r - r^2\mathbf S \right],
\end{align}
whose components are
\begin{subequations}
\begin{align}
    O_{x} &= \frac{2}{\sqrt{10}} \left[ 3xyS_{y} + 3zxS_{z} + (2x^{2} - y^{2} - z^{2})S_{x} \right] ,
\end{align}
\begin{align}
    O_{y} &= \frac{2}{\sqrt{10}} \left[ 3xyS_{x} + 3yzS_{z} + (2y^{2} - x^{2} - z^{2})S_{y} \right],
\end{align}
\begin{align}
    O_{z} &= \frac{2}{\sqrt{10}} \left[ 3zxS_{x} + 3yzS_{y} + (2z^{2} - x^{2} - y^{2})S_{z} \right].
\end{align}
\end{subequations}
In the strict irreducible-tensor sense, $\mathbf{O}$ is the rank-one AMD of the reducible Cartesian MO tensor~\cite{Kusunose20JPSJ,Sato26npjQM}. Throughout the present work, however, we refer to $\mathbf{O}$ as the MO operator in the Cartesian-multipole sense, emphasizing its microscopic structure as a spin degree of freedom coupled to an anisotropic rank-two orbital tensor.

\subsection{Atomic orbital operator representation of magnetic octupole operators}

For localized atomic states, spatial multipole operators can be represented by orbital angular momentum (OAM) operators $\mathbf{L}$. For example, we project the operator $\mathbf{O}$ onto the $p$-orbital subspace. For $l=1$, the quadratic position operators satisfy
\begin{equation}
P r_i r_j P
=
\frac{\langle r^2\rangle}{5}
\left[
3\delta_{ij}
-\frac{\{L_i,L_j\}}{\hbar^2}
\right],
\label{eq:rirj_projection}
\end{equation}
where $P$ denotes the projector onto the $p$-orbital manifold and
\begin{equation}
\{L_i,L_j\}=L_iL_j+L_jL_i
\end{equation}
is the anticommutator. In particular, for $i\neq j$,
\begin{equation}
P r_i r_j P
=
-\frac{\langle r^2\rangle}{5\hbar^2}
\{L_i,L_j\},
\end{equation}
whereas the diagonal components are
\begin{equation}
P r_i^2 P
=
\frac{\langle r^2\rangle}{5}
\left(
3-\frac{2}{\hbar^2}L_i^2
\right).
\end{equation}
Using Eq.~(\ref{eq:D_def}), the projected MO operator can be written compactly as
\begin{equation}
P O_i P
=
\frac{2\langle r^2\rangle}{5\sqrt{10}\hbar^2}
\left[
2L^2S_i
-
3\sum_j\{L_i,L_j\}S_j
\right].
\label{eq:MO_L_representation}
\end{equation}
For the $p$-orbital manifold, $L^2=l(l+1)\hbar^{2}=2\hbar^{2}$. In the following, we adopt the convention $\langle r^2\rangle=1$, so that the radial matrix element is absorbed into the definition of the MO operator. The individual components are then
\begin{align}
O_x
&=
\frac{2}{5\sqrt{10}\hbar^2}
\bigg[
2(L^2-3L_x^2)S_x
-3\{L_x,L_y\}S_y
-3\{L_x,L_z\}S_z
\bigg],
\\
O_y
&=
\frac{2}{5\sqrt{10}\hbar^2}
\bigg[
-3\{L_x,L_y\}S_x
+2(L^2-3L_y^2)S_y
-3\{L_y,L_z\}S_z
\bigg],
\\
O_z
&=
\frac{2}{5\sqrt{10}\hbar^2}
\bigg[
-3\{L_x,L_z\}S_x
-3\{L_y,L_z\}S_y
+2(L^2-3L_z^2)S_z
\bigg].
\label{eq:A18}
\end{align}

The explicit representation in Eqs.~\eqref{eq:rirj_projection}-\eqref{eq:A18} is specialized to the $p$-orbital manifold. For the $d$-orbital altermagnetic model, we use the same Cartesian MO operator in Eq.~\eqref{eq:D_def}, projected onto the $t_{2g}$ orbital subspace.

\section{General construction of magnetic multipole operators compatible with $J_{z}$}

In this section, we establish a general rule for constructing magnetic multipole operators that are compatible with a given component of the total angular momentum $J_{i}$. Without loss of generality, we consider the $i = z$ direction and set $\hbar = 1$. The tensor-product symbol, $\otimes$, is omitted for simplicity in this section.

\subsection{General compatibility rule}

Magnetic multipole operators are constructed from spatial coordinates and spin degrees of freedom. Let $\mathcal{Q}$ and $\mathcal{S}$ be operators which consist of position and spin operators, respectively. We suppose that there exist numbers $u$ and $v$ such that
\begin{align}
    [L_{z}, \mathcal{Q}] = u \mathcal{Q}, \qquad [S_{z}, \mathcal{S}] = v \mathcal{S}.
\end{align}
Since $J_{z} = L_{z} + S_{z}$, a product operator $\mathcal{Q} \mathcal{S}$ obeys
\begin{align}
    [J_{z}, \mathcal{Q} \mathcal{S}] = (u + v)\mathcal{Q} \mathcal{S}.
\end{align}
Therefore, the operator $\mathcal{Q} \mathcal{S}$ is compatible with $J_{z}$, i.e., $[J_{z}, \mathcal{Q}\mathcal{S}]=0$ if 
\begin{align}
    u + v = 0.
\end{align}
In the present work, we focus on the case in which $\mathcal{S}$ is linear in $S_{i}$ ($i \in \{x, y, z\}$). Then, $v$ is restricted to $0, \pm 1$. There are two possibilities: longitudinal magnetic multipoles ($\mathcal{S} = S_{z}$) satisfy $(u, v) = (0,0)$; transverse magnetic multipoles, constructed from $ S_{x}$ and $S_{y}$, satisfy $(u, v) = (\pm 1 , \mp 1)$, as discussed below.

\subsection{Case 1: Longitudinal magnetic multipoles}

In the longitudinal case, we have $u=v=0$ so $\mathcal{Q}$ must satisfy
\begin{align}
    [L_{z}, \mathcal{Q}] = 0.
\end{align}
Then, any operator of the form
$M^{(n)}_{z,\parallel} = \mathcal{Q}^{(n-1)}S_{z}$ commutes with $J_{z}$. Here, we note that the subscript $z$ of $M^{(n)}_{z,\parallel}$ indicates that this operator is compatible with $J_{z}$. The particular magnetic-multipole character is determined by the spatial structure of $\mathcal{Q}$.

As an example, consider a spatial multipole that is quadratic in the coordinates. The combination
\begin{align}
    \mathcal{Q}^{(2)}_{z, \parallel} = 2z^{2} - x^{2} - y^{2}
\end{align}
satisfies
\begin{align}
    [L_{z}, \mathcal{Q}^{(2)}_{z, \parallel}] = 0.
\end{align}
Consequently,
\begin{align}
    M^{(3)}_{z, \parallel} = (2z^{2} - x^{2} - y^{2}) S_{z}
\end{align}
is compatible with $J_{z}$.

\subsection{Case 2: Transverse magnetic multipoles}

In the transverse case, we have $v = \pm 1$ so we take $\mathcal{S} = S_{\pm}$ ($S_{\pm} = S_{x} \pm i S_{y}$). Then, the general compatibility rule requires the spatial part to satisfy
\begin{align}\label{eq:transverse_spatial_rule}
    [L_{z},\mathcal{Q}_{\pm}] = \pm \mathcal{Q}_{\pm}
\end{align}
It is useful to write
\begin{align}
    \mathcal{Q}_{+} &= A + iB, \qquad \mathcal{Q}_{-} = A - iB = \mathcal{Q}_{+}^{\dagger},
\end{align}
where $A$ and $B$ are Hermitian spatial operators. Then, Eq.~\eqref{eq:transverse_spatial_rule} can be written as
\begin{align}\label{eq:transverse_spatial_decomp_rule}
    [L_{z}, A] &= iB, \qquad [L_{z}, B] = -iA.
\end{align}
Thus, $A$ and $B$ form a pair that transforms as in-plane Cartesian components under rotation $L_{z}$ about the $z$ axis.

According to the general compatibility rule, $\mathcal{Q}_{+}$ must be combined with $S_{-}$ and $\mathcal{Q}_{-}$ must be combined with $S_{+}$. Therefore, a Hermitian combination of $\mathcal{Q} \mathcal{S}$ is
\begin{align}
    M^{(n)}_{z,\perp} &= \frac{\mathcal{Q}_{+}S_{-} + \mathcal{Q}_{-}S_{+}}{2} = AS_{x} + BS_{y},
\end{align}
and
\begin{align}
    A = \text{Re}[\mathcal{Q}_{+}] = \frac{\mathcal{Q}_{+}+\mathcal{Q}_{-}}{2}, \qquad B = \text{Im}[\mathcal{Q}_{+}]= \frac{\mathcal{Q}_{+}-\mathcal{Q}_{-}}{2i}.
\end{align}
Therefore, any spatial pair $(A, B)$ satisfying Eq.~\eqref{eq:transverse_spatial_decomp_rule} generates a transverse magnetic multipole compatible with $J_{z}$.

We introduce the spherical coordinates $r_{\pm} = x \pm iy$, which obey
\begin{align}
    [L_{z}, r_{\pm}] &= \pm r_{\pm}, \qquad [L_{z}, z] = 0.
\end{align}
Consider a monomial of the form
\begin{align}
    \mathcal{Q}_{n_{+}n_{-}n_{z}} = r_{+}^{n_{+}} r_{-}^{n_{-}} z^{n_{z}},
\end{align}
where $n_{+}$, $n_{-}$, and $n_{z}$ are nonnegative integers. We have
\begin{align}
    [L_{z}, \mathcal{Q}_{n_{+}n_{-}n_{z}}] = (n_{+} - n_{-}) \mathcal{Q}_{n_{+}n_{-}n_{z}}.
\end{align}
Therefore, the condition for a transverse $J_{z}$-compatible operator requires
\begin{align}\label{eq:degree_rule}
    n_{+} - n_{-} = \pm 1.
\end{align}
At a given degree $N = n - 1 = n_{+} + n_{-} + n_{z}$, Eq.~\eqref{eq:degree_rule} provides a criterion for the possible spatial structures.

For example, consider a transverse Cartesian magnetic octupole component compatible with $J_{z}$, $M^{(n=3)}_{z,\perp}$. The total polynomial degree of the spatial part is given by $N = n-1 = 2$, and the only solutions are
\begin{align}
    \mathcal{Q}_{\pm} = r_{\pm}z = (x \pm i y)z.
\end{align}
Thus,
\begin{align}
    A = xz, \qquad B = yz,
\end{align}
and the corresponding magnetic octupole operator is
\begin{align}
    M^{(3)}_{z,\perp} = xzS_{x} + yzS_{y}.
\end{align}
This provides the transverse magnetic octupole component presented in Fig.1(c) in the main text.

We also provide another example of a Cartesian magnetic triakontadipole ($n = 5$) component. In this case, the total polynomial degree of the spatial part is $N = 4$, and we have two equations
\begin{align}
    n_{+} + n_{-} + n_{z} = 4, \qquad n_{+} - n_{-} = \pm 1.
\end{align}
The possible solutions are
\begin{align}
    (A, B) = (xz^{3}, yz^{3}),
\end{align}
 and
 \begin{align}
     (A, B) = (x(x^{2}+y^{2})z, y(x^{2} + y^{2})z),
 \end{align}
which generate the two operators
\begin{align}
    xz^{3}S_{x} + yz^{3}S_{y}, \qquad x(x^{2}+y^{2})zS_{x} + y(x^{2} + y^{2})zS_{y},
\end{align}
respectively.

\subsection{Generalization to noncommuting vector operators}

The construction above can be generalized to vector operators whose Cartesian components do not commute. Let $\mathbf{V} = (V_{x}, V_{y}, V_{z})$ be a Hermitian vector operator acting in the spatial or orbital sector and transforming under rotations as
\begin{align}
    [L_{z}, V_{i}] = i\varepsilon_{zij}V_{j}.
\end{align}
The operator $\mathbf{V}$ may be either polar or axial vector, since the distinction does not affect its transformation under proper rotations. We further assume that $\mathbf{V}$ commutes with the spin operators. Introducing
\begin{align}
    V_{\pm} = V_{x} \pm iV_{y},
\end{align}
we obtain
\begin{align}
    [L_{z}, V_{\pm}] = \pm V_{\pm}, \qquad [L_{z}, V_{z}] = 0.
\end{align}
Consider an ordered product containing $n_{+}$, $n_{-}$, and $n_{z}$ factors of $V_{+}$, $V_{-}$, and $V_{z}$, respectively. Then, we have the same condition $n_{+} - n_{-} = \pm 1$, which generates the transverse component compatible with $J_{z}$. The difference from the position-operator construction arises from operator ordering. Since $[V_{i}, V_{j}] \neq 0$, different orderings of the same Cartesian factors generally represent different operators. We introduce the fully symmetrized product
\begin{align}
    \bar{\mathcal{Q}}_{n_{+}n_{-}n_{z}} = \text{Sym} [V_{+}^{n_{+}}V_{-}^{n_{-}}V_{z}^{n_{z}}].
\end{align}

For example, the noncommuting analogue of the transverse component of the magnetic octupole $M^{(3)}_{z,\perp} = xzS_{x} + yzS_{y}$ is
\begin{align}
    M^{(3)}_{z,\perp} = \frac{1}{2} \left[ \{V_{x}, V_{z}\}S_{x} + \{V_{y}, V_{z}\}S_{y} \right].
\end{align}

\section{Computational details}\label{appendix:C}

In this section, we provide the details of the two tight-binding models investigated in the main text and the calculation methods.

\subsection{Triangular lattice $p$-orbital model}

We first consider a two-dimensional triangular lattice tight-binding model of $p$ orbitals. The Bravais lattice $\mathbf{R}$ in the triangular lattice is set to $\mathbf{R} = \sum_{i=1,2}n_{i}\mathbf{a}_{i}$ $(n_{i} \in \mathbb{Z})$, where
\begin{align}
    \mathbf{a}_{1} = a \left( \frac{1}{2} \hat{\mathbf{x}} + \frac{\sqrt{3}}{2}\hat{\mathbf{y}} \right), \qquad \mathbf{a}_{2} = a \left( \frac{1}{2} \hat{\mathbf{x}} - \frac{\sqrt{3}}{2}\hat{\mathbf{y}} \right),
\end{align}
where $a$ is the lattice constant. For convenience, we also introduce the third nearest neighbor vector $\mathbf{a}_{3} = \mathbf{a}_{1} + \mathbf{a}_{2} = a\hat{\mathbf{x}}$. Each lattice site hosts the three $p$ orbitals, $p_{x}$, $p_{y}$, and $p_{z}$. We use the basis $\{\ket{p_{x}}, \ket{p_{y}}, \ket{p_{z}}\} \otimes \{\ket{\uparrow}, \ket{\downarrow}\}$.

The Hamiltonian is written as
\begin{align}
    H(\mathbf{k}) = H_{0}(\mathbf{k}) + H_{\text{soc}},
\end{align}
where $H_{0}$ contains the onsite energies, nearest-neighbor hopping, and inversion-symmetry-breaking (ISB) term as follows:
\begin{align}
    H_{0} &= H_{\rm NN} + \left[ V_{\rm ISB} \sum_{\langle ij\rangle,s} \sum_{\mu=x,y} \hat{\delta}_{ij,\mu} \left( c^{\dagger}_{i\mu s}c_{j z s} - c^{\dagger}_{i z s}c_{j\mu s} \right) + \text{h.c.} \right] + \Delta_{\text{CF}} \sum_{i,s} c^{\dagger}_{i z s}c_{i z s} + \mathcal{E}_{0}\sum_{i,s}\sum_{\mu = x,y,z}c^{\dagger}_{i\mu s}c_{i\mu s}.
\end{align}
In momentum space, it is written as
\begin{align}
    H_{0}(\mathbf{k}) = \left(
    \begin{array}{ccc}
        H_{p_{x}p_{x}} & H_{p_{x}p_{y}} & H_{p_{x}p_{z}} \\
        H_{p_{y}p_{x}} & H_{p_{y}p_{y}} & H_{p_{y}p_{z}} \\
        H_{p_{z}p_{x}} & H_{p_{z}p_{y}} & H_{p_{z}p_{z}}
    \end{array}\right)
    \otimes \sigma_{0},
\end{align}
where
\begin{subequations}\label{TB1_kinetic_components}
\begin{align}
    H_{p_{x}p_{x}} &= \mathcal{E}_{0} + \left( \frac{1}{2}t_{p\sigma} + \frac{3}{2}t_{p\pi} \right) \left[ \text{cos}(\mathbf{k} \cdot \mathbf{a}_{1}) + \text{cos}(\mathbf{k} \cdot \mathbf{a}_{2}) \right] + 2t_{p\sigma}\text{cos}(\mathbf{k} \cdot \mathbf{a}_{3}),
    \\[2mm]
    H_{p_{y}p_{y}} &= \mathcal{E}_{0} + \left( \frac{3}{2}t_{p\sigma} + \frac{1}{2}t_{p\pi} \right) \left[ \text{cos}(\mathbf{k} \cdot \mathbf{a}_{1}) + \text{cos}(\mathbf{k} \cdot \mathbf{a}_{2}) \right] + 2t_{p\pi}\text{cos}(\mathbf{k} \cdot \mathbf{a}_{3}),
    \\[2mm]
    H_{p_{z}p_{z}} &= \mathcal{E}_{0} + \Delta_{\text{CF}} + 2t_{p\pi} \sum_{i=1}^{3} \text{cos}(\mathbf{k} \cdot \mathbf{a}_{i}),
    \\[2mm]
    H_{p_{x}p_{y}} &= \frac{\sqrt{3}}{2} \left( t_{p\sigma} - t_{p\pi} \right) \left[ \text{cos}(\mathbf{k} \cdot \mathbf{a}_{1}) - \text{cos}(\mathbf{k} \cdot \mathbf{a}_{2}) \right],
    \\[2mm]
    H_{p_{x}p_{z}} &= iV_{\text{ISB}} \left[ \text{sin}(\mathbf{k} \cdot \mathbf{a}_{1}) + \text{sin}(\mathbf{k} \cdot \mathbf{a}_{2}) + 2 \, \text{sin}(\mathbf{k} \cdot \mathbf{a}_{3}) \right],
    \\[2mm]
    H_{p_{y}p_{z}} &= i\sqrt{3}V_{\text{ISB}} \left[ \text{sin}(\mathbf{k} \cdot \mathbf{a}_{1}) - \text{sin}(\mathbf{k} \cdot \mathbf{a}_{2}) \right],
\end{align}
\end{subequations}
where $V_{\text{ISB}}$ parametrizes the inversion symmetry breaking. Here, $\mathcal{E}_{0}$ denotes the common onsite energy of $p_{x}$ and $p_{y}$ orbitals, while $\Delta_{\text{CF}}$ denotes the onsite-energy difference between the $p_{z}$ and $p_{x,y}$ orbitals. The ISB hopping captures the orbital hybridization induced by surface inversion asymmetry, which underlies orbital Rashba physics in multiorbital surface states~\cite{Kim13PRb,Go17SR}. The atomic spin-orbit coupling is included as
\begin{align}
    H_{\text{soc}} = \frac{\lambda_{\text{soc}}}{\hbar^{2}} \mathbf{L} \cdot \mathbf{S},
\end{align}
where $\mathbf{L}$ and $\mathbf{S}$ are the OAM and spin operators in the $p$-orbital manifold.

This model is the tight-binding model used for the results in Fig. 3 of the main text. For the calculations presented in Fig. 3 of the main text, we use $\mathcal{E}_{0} = 0.50$ eV, $t_{p\sigma} = -0.50$ eV, $t_{p\pi} = 0.10$ eV, $\Delta_{\text{CF}} = -3.0$ eV, $V_{\text{ISB}} = -0.10$ eV, and $\lambda_{\text{soc}} = 0.40$ eV. Finally, we note that the overall common onsite energy $\mathcal{E}_{0}$, which does not affect the eigenstates, is omitted in the main text.

\subsection{Square lattice $d$-orbital altermagnetic model}\label{Appendix_C2}

Next, we consider a two-dimensional collinear altermagnetic model on a square lattice. The primitive lattice vectors are
\begin{align}
    \mathbf{a}_{1} = \frac{a}{\sqrt{2}}(1, -1, 0), \qquad \mathbf{a}_{2} = \frac{a}{\sqrt{2}}(1, 1, 0),
\end{align}
and the unit cell contains two sublattices $A$ and $B$ located at
\begin{align}
    \mathbf{r}_{A} = 0, \qquad \mathbf{r}_{B} = \frac{1}{2} (\mathbf{a}_{1} + \mathbf{a}_{2}).
\end{align}
Each sublattice hosts three $t_{2g}$ orbitals, $d_{xy}$, $d_{yz}$, and $d_{zx}$. We use the basis $\{ \ket{A}, \ket{B} \} \otimes \{ \ket{d_{xy}}, \ket{d_{yz}}, \ket{d_{zx}} \} \otimes \{ \ket{\uparrow}, \ket{\downarrow} \}$. This is the same structure as the collinear square-lattice altermagnetic model introduced in Ref.~\cite{Jo25PRL}. In this basis, the Hamiltonian can be written in the block form
\begin{align}
H(\mathbf k)= \left(
\begin{array}{cc}
    H_A(\mathbf k) & T(\mathbf k) \\
    T^{\dagger}(\mathbf k) & H_B(\mathbf k)
\end{array} \right) .
\end{align}
The inter-sublattice nearest-neighbor hopping is diagonal in both orbital and spin spaces,
\begin{align}
    T(\mathbf k) = {\rm diag} \left( h_{xy},h_{xy}, h_{yz},h_{yz}, h_{zx},h_{zx} \right),
\end{align}
where
\begin{align}
    h_{xy}=2t_{\pi}(c_x+c_y), \quad h_{yz}=2(t_{\delta}c_x+t_{\pi}c_y), \quad h_{zx}=2(t_{\pi}c_x+t_{\delta}c_y).
\end{align}
Here, $d = a/\sqrt{2}$, $c_i=\cos(k_i d)$, and $s_i=\sin(k_i d)$. These hopping amplitudes follow from the nearest-neighbor hopping of the $t_{2g}$ orbitals.

The local Hamiltonian on each sublattice contains the $sd$-exchange interaction, atomic spin-orbit coupling, sublattice-dependent crystal field splitting, and intra-sublattice ISB orbital hopping. Introducing $\eta_{A} = 1$ and $\eta_{B} = -1$, the two sublattice blocks can be expressed as
\begin{align}\label{eq:Hamiltonian_AM_eta}
    H_{\eta}(\mathbf{k}) = -\eta \frac{2J_{sd}}{\hbar} \mathbf{N} \cdot \mathbf{S} + \eta \frac{\Delta_{\text{CF}}}{2\hbar^{2}}(L^{2}_{y} - L^{2}_{x}) + \frac{\lambda_{\text{soc}}}{\hbar^{2}} \mathbf{L} \cdot \mathbf{S} + H_{\text{ISB}}^{\eta}(\mathbf{k}),
\end{align}
where $H_{A(B)} = H_{\eta = \pm 1}$ and $\mathbf{N} = \hat{\mathbf{x}}$ denotes the N\'eel-vector direction. The broken inversion symmetry effect is introduced through an intra-sublattice ISB orbital hopping:
\begin{align}
    H_{\text{ISB}}^{\eta}(\mathbf{k}) = \frac{1}{\hbar} \left[ g_{x}(\mathbf{k})L_{y} - g_{y}(\mathbf{k})L_{x} \right],
\end{align}
where
\begin{align}
    g_{x}(\mathbf{k}) = 2 \lambda_{\text{ISB}} s_{x}c_{y}, \qquad g_{y}(\mathbf{k}) = 2 \lambda_{\text{ISB}} c_{x}s_{y}.
\end{align}
Since this term has the same sign on the two magnetic sublattices, the full 12-band representation is
\begin{align}
    H_{\text{ISB}}(\mathbf{k}) = \eta_{0} \otimes H_{\text{ISB}}^{\eta}(\mathbf{k}),
\end{align}
where $\eta_{i}$ are Pauli matrices in the $A/B$ sublattice space. For $\lambda_{\text{ISB}}=0$, the model reduces to the inversion-symmetric square-lattice altermagnetic model of Ref.~\cite{Jo25PRL,Jo26arXiv}.

This model is the tight-binding model used for the results in Fig. 4 of the main text. For the calculations presented in Fig. 4 of the main text, we use $t_{\pi} = 0.18$ eV, $t_{\delta} = -0.18$ eV, $J_{sd} = 1.5$ eV, $\Delta_{\text{CF}}=0.7$ eV, $\lambda_{\text{soc}} = 0.02$ eV, and $\lambda_{\text{ISB}}=0.05$ eV.

The sublattice-resolved expectation values are evaluated using the projected operators
\begin{align}
    D_{\eta} = \frac{1}{2} \left\{P_{\eta}, D \right\},
\end{align}
where $\eta = A, B$, $P_{\eta} = \ket{\eta}\bra{\eta}$, and $D$ denotes either the spin or MO operator.

The spin and MO Rashba-Edelstein susceptibilities are evaluated using the constant relaxation-time intraband expression given in Eq. (10) of the main text. The BZ integration is performed on a $400 \times 400$ uniform $k$-mesh. The derivative of the Fermi-Dirac distribution is evaluated at $T = 300$ K.

\section{$k \cdot p$ analysis of the triangular-lattice $p$-orbital model}

In this section, we derive the low-energy Hamiltonian of the triangular-lattice $p$-orbital model introduced in Appendix~\ref{appendix:C} and clarify the origin of the distinct spin and MO textures shown in the main text. Expanding the tight-binding Hamiltonian around the $\Gamma$ point to linear order in $\mathbf{k}$, we obtain
\begin{align}\label{TB1_effective_kinetic}
    H_{0}^{\Gamma}(\mathbf{k}) = \left(
    \begin{array}{ccc}
        \mathcal{E}_{0} + \mathcal{E}_{\parallel} & 0 & i\alpha_{\text{OR}}k_{x} \\
        0 & \mathcal{E}_{0} + \mathcal{E}_{\parallel} & i\alpha_{\text{OR}}k_{y} \\
        -i\alpha_{\text{OR}} k_{x} & -i\alpha_{\text{OR}}k_{y} & \mathcal{E}_{0} + \Delta_{\text{CF}} + 6t_{p\pi}
    \end{array}\right)
    \otimes \sigma_{0},
\end{align}
where $\mathcal{E}_{\parallel} = 3(t_{p\sigma} + t_{p\pi})$ and $\alpha_{\text{OR}} = 3V_{\text{ISB}}a$. It is useful to define $\Delta_{\Gamma} = \mathcal{E}_{\parallel} - (\Delta_{\text{CF}} + 6t_{p\pi}) = 3(t_{p\sigma} - t_{p\pi}) - \Delta_{\text{CF}}$. We can decompose $H_{0}^{\Gamma}(\mathbf{k}) = H_{\text{CF}}^{\Gamma} + H_{\text{OR}}^{\Gamma}(\mathbf{k})$, where
\begin{align}\label{TB1_effective_kinetic_CF}
    H_{\text{CF}}^{\Gamma} = \left(
    \begin{array}{ccc}
        \mathcal{E}_{0} + \mathcal{E}_{\parallel} & 0 & 0 \\
        0 & \mathcal{E}_{0} + \mathcal{E}_{\parallel} & 0 \\
        0 & 0 & \mathcal{E}_{0} + \Delta_{\text{CF}} + 6t_{p\pi}
    \end{array}\right)
    \otimes \sigma_{0}
\end{align}
contains the diagonal parts due to the crystal field splitting, and
\begin{align}\label{TB1_effective_kinetic_OR}
    H_{\text{OR}}^{\Gamma}(\mathbf{k}) &= \left(
    \begin{array}{ccc}
        0 & 0 & i\alpha_{\text{OR}}k_{x} \\
        0 & 0 & i\alpha_{\text{OR}}k_{y} \\
        -i\alpha_{\text{OR}}k_{x} & -i\alpha_{\text{OR}}k_{y} & 0
    \end{array}\right)
    \otimes \sigma_{0} = \frac{\alpha_{\text{OR}}}{\hbar}\mathbf{L} \cdot (\hat{\mathbf{z}} \times \mathbf{k})
\end{align}
contains the off-diagonal parts that arise from the ISB. Together with the atomic SOC, we obtain the low-energy Hamiltonian as
\begin{align}
    H^{\Gamma} = H^{\Gamma}_{0} + \frac{\lambda_{\text{soc}}}{\hbar^{2}} \mathbf{L} \cdot \mathbf{S}.
\end{align}

At $\mathbf{k} = 0$, $J_{z} = L_{z} + S_{z}$ remains a good quantum number, while $J^{2}$ does not due to the crystal field. We introduce $\ket{p_{\pm}} = (\ket{p_{x}} \pm i\ket{p_{y}})/\sqrt{2}$. The $|J_{z}| = 3/2$ doublet,
\begin{align}
    \left\vert J_{z} = +\frac{3}{2} \right\rangle = \ket{p_{+}, \uparrow}, \qquad \left\vert J_{z} = -\frac{3}{2} \right\rangle = \ket{p_{-}, \downarrow},
\end{align}
does not hybridize with the $p_{z}$ states and has the energy
\begin{align}
    \mathcal{E}_{3/2} = \mathcal{E}_{0} + \mathcal{E}_{\parallel} + \frac{\lambda_{\text{soc}}}{2}.
\end{align}
Meanwhile, $J_{z} = \pm 1/2$ sector contains both $p_{x,y}$ state and $p_{z}$ state. In the basis $\{ \ket{p_{+},\downarrow}, \ket{p_{z},\uparrow} \} (\{ \ket{p_{-},\uparrow}, \ket{p_{z},\downarrow} \})$, the $J_{z} = \pm 1/2$ blocks are written as
\begin{align}
    H_{J_{z} = \pm 1/2} = \left(
    \begin{array}{cc}
        \mathcal{E}_{0} + \mathcal{E}_{\parallel} - \frac{\lambda_{\text{soc}}}{2} & \mp \frac{\lambda_{\text{soc}}}{\sqrt{2}} \\
        \mp \frac{\lambda_{\text{soc}}}{\sqrt{2}} & \mathcal{E}_{0} + \Delta_{\text{CF}} + 6t_{p\pi}
    \end{array}\right) ,
\end{align}
respectively. We take
\begin{align}
    \delta = \Delta_{\Gamma} - \frac{\lambda_{\text{soc}}}{2}, \quad \Omega = \sqrt{\delta^{2} + 2\lambda_{\text{soc}}^{2}},
\end{align}
and a mixing angle through
\begin{align}
    \text{cos} \, 2\theta = \frac{\delta}{\Omega}, \qquad \text{sin} \, 2\theta = \frac{\sqrt{2}\lambda_{\text{soc}}}{\Omega}.
\end{align}
For $\Delta_{\Gamma} > 0$, the upper $|J_{z}| = 1/2$ doublet is predominantly $p_{x,y}$, while the lower doublet is predominantly $p_{z}$. Their energies are
\begin{align}
    \mathcal{E}_{U,L} = \mathcal{E}_{0} + \frac{2\mathcal{E}_{\parallel} - \lambda_{\text{soc}} + 2\Delta_{\text{CF}} + 12t_{p\pi}}{4} \pm \frac{\Omega}{2},
\end{align}
respectively. This identifies the three Kramers-doublet sectors corresponding to the three band groups discussed in the main text. Throughout the main text, we refer to the predominantly $p_{x,y}$ $|J_{z}| = 1/2$ doublet simply as the ``$|J_{z}| = 1/2$ sector", the other $|J_{z}| = 1/2$ doublet as the ``predominantly $p_{z}$ sector", and the remaining unmixed doublet as the $|J_{z}| = 3/2$ sector, respectively.

Now, we investigate the Hamiltonian around the $\Gamma$ point. Since $H_{\text{OR}}$ is linear in $\mathbf{k}$, it makes no contribution at exactly $\Gamma$, but should be considered near $\Gamma$ to obtain the total Hamiltonian. We choose the states of the $|J_{z}| = 1/2$ sector as
\begin{subequations}\label{eq:admixture}
\begin{align}
    \ket{U, +} &= \text{cos} \, \theta \, \ket{p_{+}, \downarrow} - \text{sin} \, \theta \, \ket{p_{z}, \uparrow}
\end{align}
\begin{align}
    \ket{U, -} &= \text{cos} \, \theta \, \ket{p_{-}, \uparrow} + \text{sin} \, \theta \, \ket{p_{z}, \downarrow}.
\end{align}
\end{subequations}
Then, the Hamiltonian is written as
\begin{align}\label{eq:MO_sector1}
    H_{U}(\mathbf{k}) = \mathcal{E}_{U} + \alpha_{\text{R}}(\mathbf{k} \times \hat{\mathbf{z}}) \cdot \bm{\sigma},
\end{align}
where $\bm{\sigma}$ denotes the pseudospin in the $\{ \ket{U,+}, \ket{U,-} \}$ space and
\begin{align}
    \alpha_{\text{R}} &= \sqrt{2} \alpha_{\text{OR}} \, \text{cos} \, \theta \, \text{sin} \, \theta = \frac{\alpha_{\text{OR}}\lambda_{\text{soc}}}{\Omega}.
\end{align}
The Hamiltonian of the predominantly $p_{z}$ sector is obtained in a similar way,
\begin{align}\label{eq:hamiltonian_sector2}
    H_{L}(\mathbf{k}) = \mathcal{E}_{L} + \alpha_{\text{R}}^{L}(\mathbf{k} \times \hat{\mathbf{z}}) \cdot \bm{\rho},
\end{align}
where $\bm{\rho}$ denotes the pseudospin in the $\{ \ket{L,+}, \ket{L,-} \}$ space. In the crystal-field dominated limit, i.e., $\Delta_{\Gamma} \gg |\lambda_{\text{soc}}|$, both $\alpha_{\text{R}}$ and $\alpha_{\text{R}}^{L}$ reduce to
\begin{align}
    \alpha_{\text{R}} = \alpha_{\text{R}}^{L} = \frac{\alpha_{\text{OR}}\lambda_{\text{soc}}}{\Delta_{\Gamma}}.
\end{align}
Thus, the Rashba splitting requires both the ISB orbital Rashba coupling and SOC. On the other hand, the $|J_{z}| = 3/2$ sector has no matrix element of $H_{\text{OR}}$ to linear order in $\mathbf{k}$. Therefore, no linear Rashba texture arises in this sector.

The distinct spin and MO textures of the three sectors are clearly exposed by projecting the physical operators onto the corresponding low-energy subspaces. We first assume the crystal-field-dominated limit. Consider the $|J_{z}| = 1/2$ sector. We take the projection operator $P_{1/2} = \ket{p_{+}, \downarrow}\bra{p_{+}, \downarrow} + \ket{p_{-}, \uparrow}\bra{p_{-}, \uparrow}$. Within this subspace, the in-plane spin and OAM operators vanish in the leading order,
\begin{align}
    P_{1/2}S_{i}P_{1/2} = P_{1/2}L_{i}P_{1/2} = 0,
\end{align}
where $i \in \{x,y\}$. It is particularly transparent in the basis $\{ \ket{p_{+}, \downarrow}, \ket{p_{-}, \uparrow} \}$. The spin operators $S_{x}$ and $S_{y}$ flip the spin without changing the orbital state, while $L_{x}$ and $L_{y}$ change the OAM without flipping the spin. Neither operation therefore connects $\ket{p_{+}, \downarrow}$ and $\ket{p_{-}, \uparrow}$.

In contrast, the MO operator does not vanish in this subspace. Using the $p$-orbital representation of $\mathbf{O}$ derived in Appendix~\ref{sec:Atomic_MO_operators}, we obtain
\begin{align}\label{eq:MO_sector2}
    P_{1/2}O_{x}P_{1/2} = O_{0}\sigma_{x}, \qquad P_{1/2}O_{y}P_{1/2} = O_{0}\sigma_{y},
\end{align}
where $O_{0} = 6\hbar / 5\sqrt{10}$. The finite matrix element of $\mathbf{O}$ originates from its spin-orbital combined structure. Unlike either $\mathbf{S}$ or $\mathbf{L}$ alone, the MO operator can simultaneously connect the different orbital and spin components of the two pseudospin states.

Combining Eq.~\eqref{eq:MO_sector2} with the effective Hamiltonian [Eq.~\eqref{eq:MO_sector1}], we find
\begin{align}
    \braket{\bm{\sigma}_{\parallel}}_{\nu} = \nu \frac{\mathbf{k} \times \hat{\mathbf{z}}}{|\mathbf{k}|},
\end{align}
where $\nu = \pm 1$. It follows that
\begin{align}
    \braket{\mathbf{O}_{\parallel}}_{\nu} = \nu O_{0} \frac{\mathbf{k} \times \hat{\mathbf{z}}}{|\mathbf{k}|}, \qquad \braket{\mathbf{S}_{\parallel}}_{\nu} = \braket{\mathbf{L}_{\parallel}}_{\nu} = 0.
\end{align}
Therefore, the Rashba pseudospin remains fully encoded in the MO channel even though its conventional in-plane magnetic-dipole components vanish. Consequently, the Hamiltonian in the $|J_{z}| = 1/2$ sector is given by
\begin{align}
    H_{U} = \mathcal{E}_{U} + \frac{\alpha_{R}}{O_{0}} (\mathbf{k} \times \hat{\mathbf{z}}) \cdot P_{1/2} \mathbf{O} P_{1/2}.
\end{align}
This result provides the microscopic origin of the pure-MO Rashba state discussed in the main text.

The predominantly $p_{z}$ sector behaves differently. We assume the same crystal-field-dominated limit and use the basis $\{ \ket{p_{z}, \uparrow}, \ket{p_{z}, \downarrow} \}$. Then, we obtain
\begin{align}
    P_{z}S_{i}P_{z} = \frac{\hbar}{2}\rho_{i}, \qquad P_{z}L_{i}P_{z} = 0, \qquad P_{z}O_{i}P_{z} = O_{0}^{L}\rho_{i},
\end{align}
where $P_{z} = \ket{p_{z}, \uparrow}\bra{p_{z}, \uparrow} + \ket{p_{z}, \downarrow}\bra{p_{z}, \downarrow}$ and $O_{0}^{L} = -2\hbar/5\sqrt{10}$. It follows that
\begin{align}
    \braket{\mathbf{O}_{\parallel}}_{\nu} = \nu O_{0}^{L} \frac{\mathbf{k} \times \hat{\mathbf{z}}}{|\mathbf{k}|}, \qquad \braket{\mathbf{S}_{\parallel}}_{\nu} = \nu \frac{\hbar}{2} \frac{\mathbf{k} \times \hat{\mathbf{z}}}{|\mathbf{k}|}.
\end{align}
This explains why the predominantly $p_{z}$ bands in the full tight-binding calculation exhibit simultaneous spin and MO Rashba textures, as discussed in the main text.

We finally consider the $|J_{z}| = 3/2$ sector. The in-plane physical operators projected onto this subspace satisfy
\begin{align}
    P_{3/2}S_{i}P_{3/2} = P_{3/2}L_{i}P_{3/2} = P_{3/2}O_{i}P_{3/2} = 0,
\end{align}
where $P_{3/2} = \ket{p_{+}, \uparrow}\bra{p_{+}, \uparrow} + \ket{p_{-}, \downarrow}\bra{p_{-}, \downarrow}$. Moreover, as shown above, $H_{\text{OR}}$ does not have a matrix element within this sector to linear order in $\mathbf{k}$. Consequently, the $|J_{z}| = 3/2$ sector has neither a linear Rashba splitting nor a leading in-plane spin or MO texture.

The exact $|J_{z}| = 1/2$ sector at finite SOC contains a small $p_{z}$ admixture, as shown in Eq.~\eqref{eq:admixture}. Projecting the spin and MO operators onto this exact doublet gives
\begin{align}
    P_{U}S_{i}P_{U} = -\frac{\hbar}{2}\text{sin}^{2} \theta \, \sigma_{i}, \qquad P_{U}L_{i}P_{U} = -\sqrt{2}\hbar \, \text{cos} \, \theta \, \text{sin} \, \theta \, \sigma_{i}, \qquad P_{U}O_{i}P_{U} = O_{U}(\theta)\sigma_{i},
\end{align}
where
\begin{align}
    O_{U}(\theta) = \frac{\hbar}{5\sqrt{10}} \left( 6\, \text{cos}^{2}\theta - 3\sqrt{2} \text{cos} \, \theta \, \text{sin} \, \theta + 2 \, \text{sin}^{2} \theta \right).
\end{align}
In the crystal-field-dominated regime, we have
\begin{align}
    \text{sin} \, \theta \approx \frac{\lambda_{\text{soc}}}{\sqrt{2} \Delta_{\Gamma}}, \qquad \text{cos} \, \theta \approx 1.
\end{align}
This implies that
\begin{align}
    S_{\parallel} = \mathcal{O} \left[ \left( \frac{\lambda_{\text{soc}}}{\Delta_{\Gamma}} \right)^{2} \right], \qquad L_{\parallel} = \mathcal{O}\left( \frac{\lambda_{\text{soc}}}{\Delta_{\Gamma}} \right), \qquad O_{\parallel} = \mathcal{O}(1).
\end{align}
Therefore, a finite SOC weakly restores the conventional spin and OAM components, while the MO component remains finite already at zeroth order in $\lambda_{\text{soc}}/\Delta_{\Gamma}$. This hierarchy explains why the full tight-binding calculation shows a strongly suppressed spin texture but a pronounced MO texture in the $|J_{z}| = 1/2$ sector.

\section{N\'eel magnetic-octupole torque in the $d$-wave altermagnetic model}

In this section, we present additional analysis for the square lattice $d$-orbital altermagnetic model introduced in the End Matter in the main text and Appendix~\ref{Appendix_C2}.

\subsection{Crystal-field dependence of the magnetic-octupole and staggered-spin responses}
\begin{figure}[h!]
\includegraphics[width=0.4\textwidth]{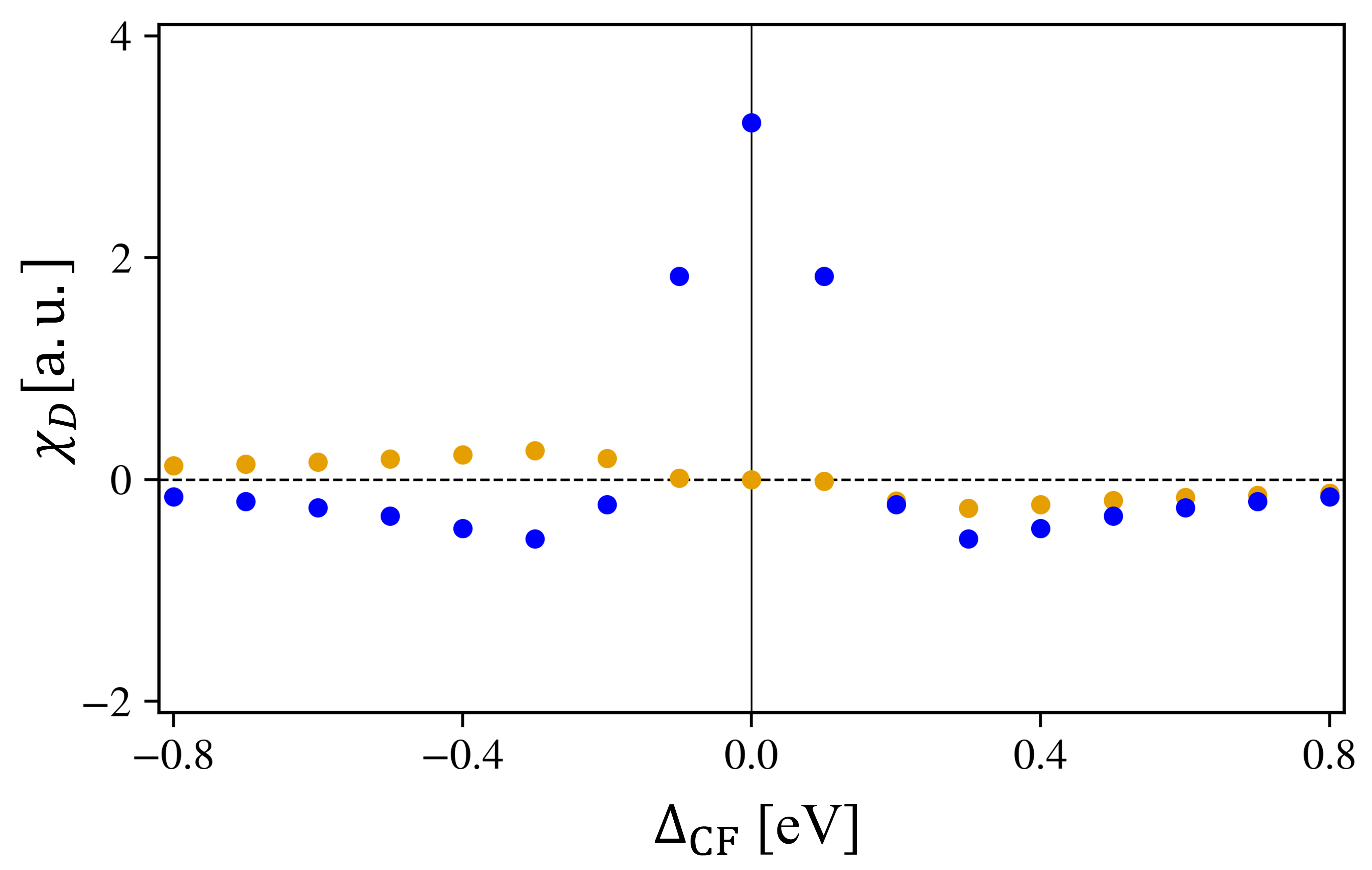}
\caption{\label{fig.s1}
Crystal-field dependence of the MO and staggered-spin Rashba-Edelstein responses in the $d$-wave altermagnetic model. The total MO susceptibility $\chi_{O, \text{tot}}$ (blue) and staggered-spin susceptibility $\chi_{S,A-B} = \chi_{S,A} - \chi_{S,B}$ (orange) are shown as functions of $\Delta_{\text{CF}}$ at $\mathcal{E} = -1.56$ eV. All other parameters are fixed to those used in Fig. 4 of the main text.}
\end{figure}

We first investigate the dependence of the MO and staggered-spin Rashba-Edelstein responses on the staggered crystal-field splitting $\Delta_{\text{CF}}$. In the altermagnetic model, the crystal-field term is odd under exchange of the two magnetic sublattices,
\begin{align}
    H_{\text{CF}} = \eta_{z} \otimes \frac{\Delta_{\text{CF}}}{2\hbar^{2}} \left( L_{y}^{2} - L_{x}^{2} \right) ,
\end{align}
where $\eta_{i}$ denotes the pseudospin for the $A/B$-sublattice. The Hamiltonians with opposite signs of $\Delta_{\text{CF}}$ are related by the combined sublattice-exchange and time-reversal operation $\eta_{x} \mathcal{T}$, where $\mathcal{T}$ is the time-reversal operator. Then, the Hamiltonian with the $\eta_{0}$-type ISB coupling used in the main text satisfies
\begin{align}
    \eta_{x}\mathcal{T}: \, H(\mathbf{k}, \Delta_{\text{CF}}) \longrightarrow H(-\mathbf{k}, -\Delta_{\text{CF}}).
\end{align}
This relation imposes distinct constraints on the MO and staggered-spin Rashba-Edelstein responses. The MO operator, $O_{y}$, is odd under time reversal and even under sublattice exchange, so it is odd under $\eta_{x}\mathcal{T}$. In contrast, the staggered spin operator, $\eta_{z}S_{y}$, is odd under both $\eta_{x}$ and $\mathcal{T}$, so it is even under $\eta_{x} \mathcal{T}$. Since the group velocity $v_{x}$ changes sign under time reversal, the constant-relaxation-time response formula gives
\begin{align}\label{eq:appendix_E3}
    \chi_{O,\text{tot}}(\Delta_{\text{CF}}) = \chi_{O,\text{tot}}(-\Delta_{\text{CF}})
\end{align}
and
\begin{align}\label{eq:appendix_E4}
    \chi_{S,A-B}(\Delta_{\text{CF}}) = -\chi_{S,A-B}(-\Delta_{\text{CF}}).
\end{align}
In particular, we have
\begin{align}\label{eq:appendix_E5}
    \chi_{S,A-B}(\Delta_{\text{CF}}=0) = 0.
\end{align}

To examine these relations, we calculate $\chi_{O,\text{tot}}$ and $\chi_{S,A-B}$ as functions of $\Delta_{\text{CF}}$ at a representative energy $\mathcal{E} = -1.56$ eV using the same parameters as in Fig. 4 of the main text except for $\Delta_{\text{CF}}$. As shown in Fig.~\ref{fig.s1}, $\chi_{O,\text{tot}}$ is even under reversal of $\Delta_{\text{CF}}$ and remains finite at $\Delta_{\text{CF}} = 0$. In contrast, $\chi_{S,A-B}$ is odd under reversal of $\Delta_{\text{CF}}$ and vanishes at $\Delta_{\text{CF}} = 0$. These results indicate that the MO Rashba-Edelstein polarization is converted into the staggered spin response through the sublattice-dependent orbital structure of the $d$-wave AM.

\subsection{Robustness against band sector mixing}

\begin{figure}[h!]
\includegraphics[width=0.5\textwidth]{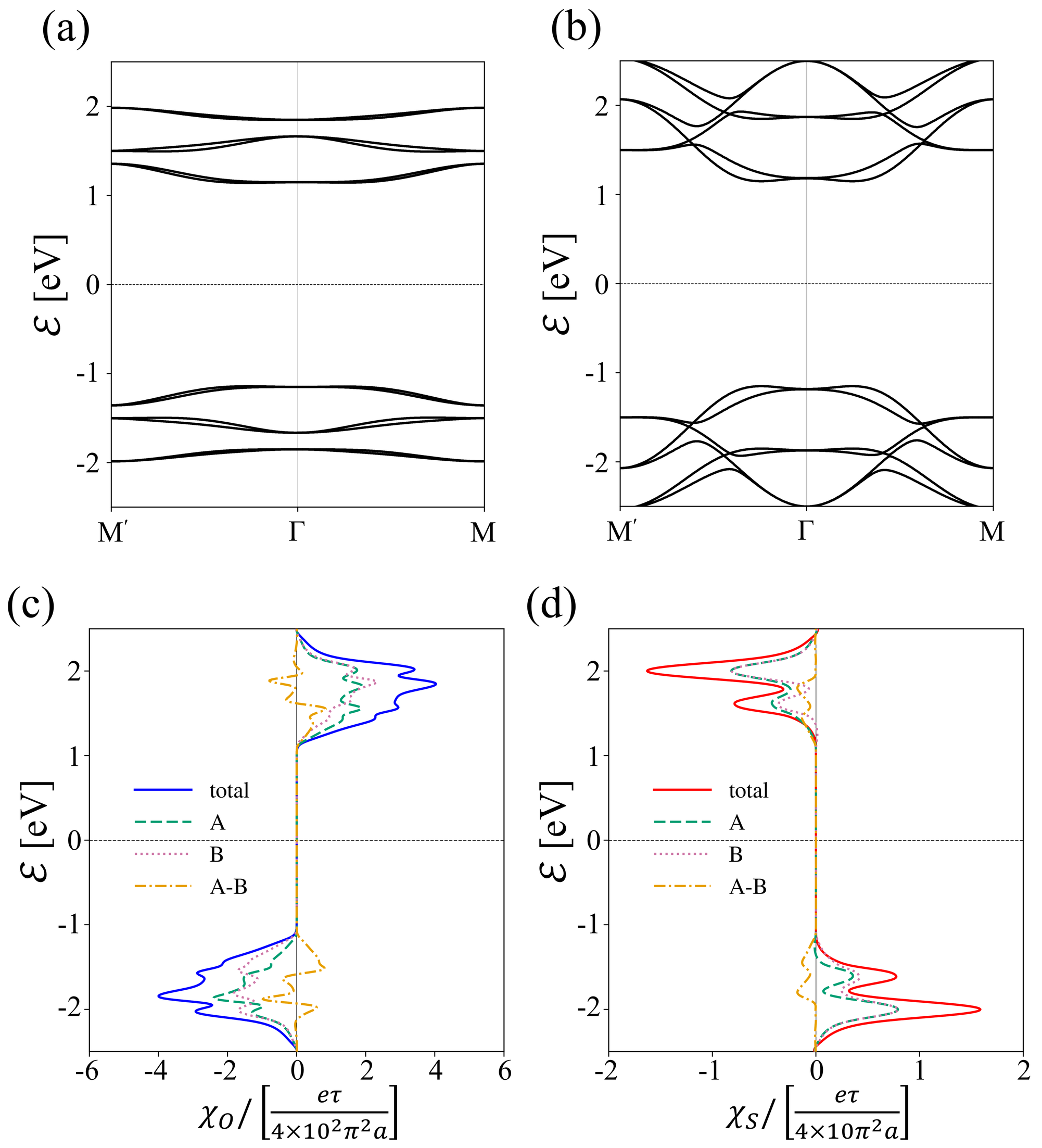}
\caption{\label{fig.s2}
(a) Band structure of the parameter set used in Fig. 4 of the main text. (b) Band structure of the parameter set presented in Appendix E.2. (c), (d) Total and sublattice-resolved MO and spin Rashba-Edelstein susceptibilities, respectively.}
\end{figure}

The parameter set used in Fig. 4 of the main text produces relatively well-separated band sectors [Fig.~\ref{fig.s2}(a)]. Consequently, the corresponding MO and spin Edelstein responses exhibit several pronounced peaks. To verify that the MO-induced staggered-spin response does not rely on this nearly isolated-sector regime, we additionally consider a more dispersive parameter set for which the bandwidth is increased and the different orbital sectors substantially overlap.

To investigate the effect of band-sector mixing, we increase the inter-sublattice hopping amplitudes while keeping the SOC and ISB strength unchanged. Specifically, we use $t_{\pi} = 0.50$ eV, $t_{\delta} = -0.36$ eV, while retaining $J_{sd} = 1.5$ eV, $\Delta_{\text{CF}} = 0.70$ eV, $\lambda_{\text{soc}} = 0.02$ eV, and $\lambda_{\text{ISB}} = 0.05$ eV as in Fig. 4 of the main text. Figure~\ref{fig.s2}(b) shows the resulting band structure. In this situation, the individual orbital sectors are no longer clearly separated. Nevertheless, the electric field still generates a finite MO Rashba-Edelstein response and a concomitant staggered-spin response [Figs.~\ref{fig.s2}(c) and \ref{fig.s2}(d)]. These results demonstrate that their emergence does not require a separation of nearly flat band sectors. Although the detailed energy dependence and the positions of the pronounced peaks are model dependent, the underlying mechanism remains unchanged.

\subsection{Additional discussion on inversion symmetry breaking terms}

In the main text, inversion symmetry breaking (ISB) is introduced by an intra-sublattice orbital hopping having the same polarity on the two magnetic sublattices. Here, we discuss alternative microscopic forms of the ISB term. The intra-sublattice ISB orbital hopping used in the main text and Appendix~\ref{Appendix_C2} is represented by
\begin{align}
    H_{\text{ISB}}(\mathbf{k}) = \frac{1}{\hbar} \eta_{0} \otimes \left[ g_{x}(\mathbf{k}) L_{y} - g_{y}(\mathbf{k})L_{x} \right],
\end{align}
where
\begin{align}
    g_{x}(\mathbf{k}) = 2\lambda_{\text{ISB}}s_{x} c_{y}, \qquad g_{y}(\mathbf{k}) = 2\lambda_{\text{ISB}}c_{x}s_{y}.
\end{align}
We generalize this to the $\eta_{i}$-type ISB as
\begin{align}
    H_{\text{ISB}}^{(i)}(\mathbf{k}) = \frac{1}{\hbar} \eta_{i} \otimes \left[ g_{x}^{(i)}(\mathbf{k}) L_{y} - g_{y}^{(i)}(\mathbf{k})L_{x} \right],
\end{align}
with $i = 0, x, z$. Here, $g^{(0)}_{x(y)} = 2\lambda_{\text{ISB}}^{(0)}s_{x(y)}c_{y(x)}$, $g^{(x)}_{x(y)} = 2\lambda_{\text{ISB}}^{(x)}s_{x(y)}$, and $g^{(z)}_{x(y)} = 2\lambda_{\text{ISB}}^{(z)}s_{x(y)}c_{y(x)}$, respectively. In the notation used in the main text and Appendix~\ref{Appendix_C2}, $H_{\text{ISB}} = H_{\text{ISB}}^{(0)}$, $g_{x,y} = g_{x,y}^{(0)}$, and $\lambda_{\text{ISB}} = \lambda_{\text{ISB}}^{(0)}$.
\begin{figure}[t!]
\includegraphics[width=0.9\textwidth]{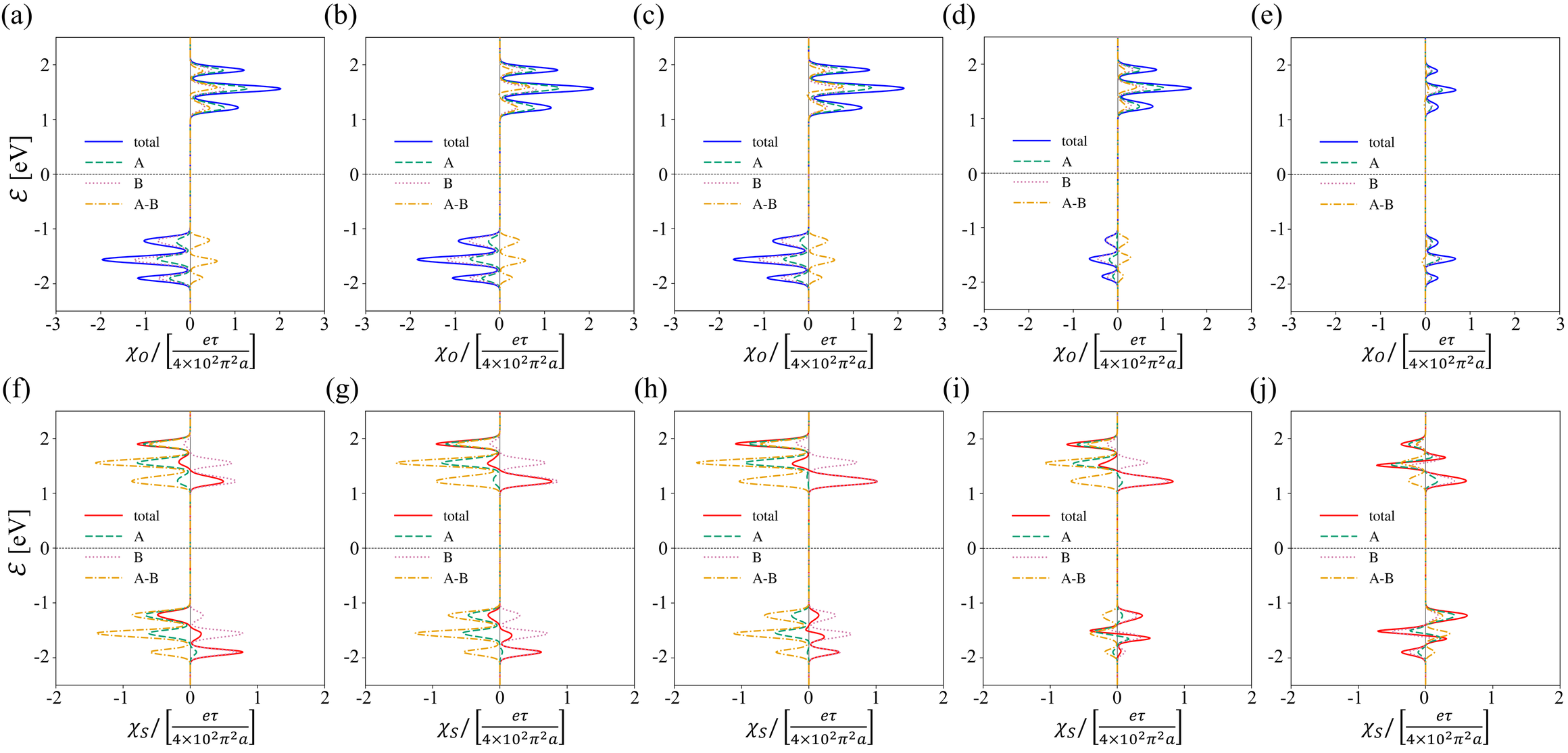}
\caption{\label{fig.s3}
The upper row (a)-(e) shows the total and sublattice-resolved MO susceptibilities, and the lower row (f)-(j) shows the corresponding spin susceptibilities. From left to right, the columns correspond to $(\lambda^{(0)}_{\text{ISB}}, \lambda^{(x)}_{\text{ISB}}) = (0.05,0.0)$, $(0.05,0.025)$, $(0.05,0.05)$, $(0.025,0.05)$, and $(0.0,0.05)$ eV, where $\lambda^{(0)}_{\text{ISB}}$ and $\lambda^{(x)}_{\text{ISB}}$ denote the strengths of the $\eta_{0}$- and $\eta_{x}$-type ISB couplings, respectively. All other parameters are fixed to those used in Fig. 4 of the main text.}
\end{figure}

The $\eta_{0}$-type describes same-sign intra-sublattice ISB hopping. It is expected when the two magnetic sublattices experience the same surface or interface polarity. The $\eta_{x}$-type term describes an inter-sublattice ISB hopping, corresponding to inversion-asymmetric orbital hybridization along hopping between the two sublattices. The $\eta_{z}$ term describes opposite local ISB polarities on the two sublattices. These different sublattice structures can arise from microscopic inversion-broken environments. For example, when a magnetic layer is placed on a substrate, capped by a different material, or subjected to a perpendicular electrostatic field, the structural asymmetry defines a common polar axis for the two magnetic sublattices. Such a global ISB can modify both intra-sublattice and inter-sublattice orbital hybridizations. This generally allows both the $\eta_{0}$- and $\eta_{x}$-type ISB terms. Their relative strengths depend on the microscopic orbital overlaps and the hopping geometry. On the other hand, the $\eta_{z}$-type term corresponds to opposite local polar environments on the two sublattices. It may occur when the two magnetic sites experience inversion-related ligand distortions or opposite local electric fields.

In the main text, we employ the $\eta_{0}$-type ISB term because it provides the clearest realization of the mechanism. Since the $\eta_{0}$-type ISB term does not directly mix the two sublattices, the generation of the orbital-dependent MO polarization can be clearly distinguished from its subsequent conversion into a staggered spin density by the sublattice-dependent orbital character. The $\eta_{x}$-type ISB also supports the same mechanism, but introduces additional inter-sublattice hybridization. This makes the separation of these two processes less transparent for the representative parameter set.

To illustrate this, Fig.~\ref{fig.s3} presents a sequence connecting the pure $\eta_{0}$-type and pure $\eta_{x}$-type limits by varying the two coupling strengths $\lambda_{\text{ISB}}^{(0)}$ and $\lambda_{\text{ISB}}^{(x)}$. Both $H_{\text{ISB}}^{(0)}$ and $H_{\text{ISB}}^{(x)}$ transform identically under the operation $\eta_{x}\mathcal{T}$. Thus, the parity relations in Eqs.~\eqref{eq:appendix_E3}-\eqref{eq:appendix_E5} remain valid for a purely inter-sublattice $\eta_{x}$-type ISB coupling as well as for coexistence of the $\eta_{0}$-and $\eta_{x}$-type terms [Fig.~\ref{fig.s3}]. The $\eta_{z}$-type ISB presents a different situation because the ISB field itself is sublattice odd. Thus, we distinguish this staggered-ISB channel from the uniform-ISB mechanism considered here.

\subsection{Current-induced N\'eel torque}

The current-induced torque is evaluated following the standard torkance formalism~\cite{Freimuth14PRb}. For a Hamiltonian depending on the N\'eel vector direction $\mathbf{N}$, we define the torque acting on the N\'eel order as
\begin{align}
    \hat{\mathbf{T}}_{N} = -\mathbf{N} \times \frac{\partial H}{\partial \mathbf{N}}.
\end{align}
For the $sd$ exchange term in Eq.~\eqref{eq:Hamiltonian_AM_eta}, this gives
\begin{align}
    \hat{\mathbf{T}}_{N} = \frac{2J_{sd}}{\hbar} \mathbf{N} \times (\eta_{z} \otimes \mathbf{S}).
\end{align}
Within the same constant-relaxation-time intraband treatment used for the Rashba-Edelstein responses, the corresponding torkance is evaluated by
\begin{align}
    \frac{T_{i}}{E_{j}} = -e\tau \sum_{n\mathbf{k}}\left( -\frac{\partial f}{\partial \mathcal{E}} \right) \braket{T_{i}}_{n\mathbf{k}} v_{n\mathbf{k},j}.
\end{align}
For the configuration used in the main text, $\mathbf{N} \parallel \mathbf{E} \parallel \hat{\mathbf{x}}$, we evaluate the $zx$ component of the torkance, $T_{z}/E_{x}$.

\section{Magnetic-octupole Rashba states in Bi$_2$Te$_3$ and Bi$_2$Se$_3$ surface states}

Following the spin-orbital surface-state wavefunctions of the Bi$_{2}$Se$_{3}$ family derived in Ref.~\cite{Zhang13PRL}, we consider the surface-state wavefunctions
\begin{subequations}
\begin{align}
|\Phi_{+}\rangle
=
\sum_{\alpha}
\Big[
&a_{\alpha}^{+}|p_z,\uparrow_{\theta}\rangle
-\frac{i}{\sqrt{2}}r_{\alpha}^{+}|p_r,\uparrow_{\theta}\rangle
+\frac{1}{\sqrt{2}}t_{\alpha}^{+}|p_t,\downarrow_{\theta}\rangle
\Big],
\end{align}
\begin{align}
|\Phi_{-}\rangle
=
\sum_{\alpha}
\Big[
&a_{\alpha}^{-}|p_z,\downarrow_{\theta}\rangle
+\frac{i}{\sqrt{2}}r_{\alpha}^{-}|p_r,\downarrow_{\theta}\rangle
-\frac{1}{\sqrt{2}}t_{\alpha}^{-}|p_t,\uparrow_{\theta}\rangle
\Big],
\end{align}
\end{subequations}
where
\begin{align}
a_{\alpha}^{+} &= u_{0,\alpha}-v_{1,\alpha}k,
&
r_{\alpha}^{+} &= v_{0,\alpha}-u_{1,\alpha}k-w_{1,\alpha}k,
&
t_{\alpha}^{+} &= v_{0,\alpha}-u_{1,\alpha}k+w_{1,\alpha}k,\notag
\\
a_{\alpha}^{-} &= u_{0,\alpha}+v_{1,\alpha}k,
&
r_{\alpha}^{-} &= v_{0,\alpha}+u_{1,\alpha}k+w_{1,\alpha}k,
&
t_{\alpha}^{-} &= v_{0,\alpha}+u_{1,\alpha}k-w_{1,\alpha}k,
\end{align}
and
\begin{subequations}
\begin{align}
    \ket{p_{r}} &= \text{cos} \, \theta_{k} \ket{p_{x}} + \text{sin} \, \theta_{k} \ket{p_{y}},
\end{align}
\begin{align}
    \ket{p_{t}} &= - \text{sin} \, \theta_{k} \ket{p_{x}} + \text{cos} \, \theta_{k} \ket{p_{y}}.
\end{align}
\end{subequations}
All coefficients $u_{0,\alpha}$, $v_{0,\alpha}$,
$u_{1,\alpha}$, $v_{1,\alpha}$, and $w_{1,\alpha}$ are taken to be real. The helical spin states are defined as
\begin{subequations}
\begin{align}
\ket{\uparrow_{\theta}}
&=
\frac{1}{\sqrt{2}}
\left(
i e^{-i\theta_k}\ket{\uparrow}
+
\ket{\downarrow}
\right),
\end{align}
\begin{align}
\ket{\downarrow_{\theta}}
&=
\frac{1}{\sqrt{2}}
\left(
- i e^{-i\theta_k}\ket{\uparrow}
+
\ket{\downarrow}
\right),
\end{align}
\end{subequations}
We then calculate the expectation values of the following operators,
\begin{align}
O_x&= \frac{2}{5\sqrt{10}\hbar^2}
\bigg[
2(L^2-3L_x^2)S_x
-3\{L_x,L_y\}S_y
-3\{L_x,L_z\}S_z
\bigg],
\\[2mm]
O_y&=\frac{2}{5\sqrt{10}\hbar^2}
\bigg[
-3\{L_x,L_y\}S_x
+2(L^2-3L_y^2)S_y
-3\{L_y,L_z\}S_z
\bigg].
\end{align}
For each orbital index $\alpha$, define
\begin{align}
F_{\pm,\alpha}(k)
=
&a_{\pm,\alpha}^{2}
-\frac{3}{\sqrt{2}}
a_{\pm,\alpha}t_{\pm,\alpha}
+\frac{1}{2}r_{\pm,\alpha}^{2}
+\frac{3}{2}r_{\pm,\alpha}t_{\pm,\alpha}
+t_{\pm,\alpha}^{2}.
\end{align}
Then
\begin{align}
\mathcal{F}_{\pm}(k)
&=
\sum_{\alpha}F_{\pm,\alpha}(k).
\end{align}

The unnormalized expectation values are
\begin{subequations}
\begin{align}
\langle\Phi_+|O_x|\Phi_+\rangle
&=-\frac{2\hbar}{5\sqrt{10}}\mathcal{F}_+(k)\sin\theta_k,
\end{align}
\begin{align}
\langle\Phi_+|O_y|\Phi_+\rangle
&=\frac{2\hbar}{5\sqrt{10}}\mathcal{F}_+(k)\cos\theta_k,
\end{align}
\begin{align}
\langle\Phi_-|O_x|\Phi_-\rangle
&=
\frac{2\hbar}{5\sqrt{10}}\mathcal{F}_-(k)\sin\theta_k,
\end{align}
\begin{align}
\langle\Phi_-|O_y|\Phi_-\rangle
&=
-\frac{2\hbar}{5\sqrt{10}}\mathcal{F}_-(k)\cos\theta_k.
\end{align}
\end{subequations}

The normalization factors are
\begin{align}
\mathcal{N}_{\pm}(k)
&=
\langle\Phi_\pm|\Phi_\pm\rangle =
\sum_{\alpha}
\left[
a_{\pm,\alpha}^{2}
+\frac{1}{2}r_{\pm,\alpha}^{2}
+\frac{1}{2}t_{\pm,\alpha}^{2}
\right].
\end{align}

Therefore, the normalized expectation values are
\begin{subequations}
\begin{align}
\langle O_x\rangle_+
&=
-\frac{2\hbar}{5\sqrt{10}}
\frac{\mathcal{F}_+(k)}{\mathcal{N}_+(k)}
\sin\theta_k,
\end{align}
\begin{align}
\langle O_y\rangle_+
&=
\frac{2\hbar}{5\sqrt{10}}
\frac{\mathcal{F}_+(k)}{\mathcal{N}_+(k)}
\cos\theta_k,
\end{align}
\begin{align}
\langle O_x\rangle_-
&=
\frac{2\hbar}{5\sqrt{10}}
\frac{\mathcal{F}_-(k)}{\mathcal{N}_-(k)}
\sin\theta_k,
\end{align}
\begin{align}
\langle O_y\rangle_-
&=
-\frac{2\hbar}{5\sqrt{10}}
\frac{\mathcal{F}_-(k)}{\mathcal{N}_-(k)}
\cos\theta_k.
\end{align}
\end{subequations}

Equivalently,
\begin{align}
\langle\mathbf{O}_{\parallel}\rangle_+
=
-\frac{2\hbar}{5\sqrt{10}}
\frac{\mathcal{F}_+(k)}{\mathcal{N}_+(k)}
\left(
\sin\theta_k,
-\cos\theta_k
\right),
\\
\langle\mathbf{O}_{\parallel}\rangle_-
=
-\frac{2\hbar}{5\sqrt{10}}
\frac{\mathcal{F}_-(k)}{\mathcal{N}_-(k)}
\left(
-\sin\theta_k,
\cos\theta_k
\right).
\end{align}

These two results can be compactly written as
\begin{equation}
\langle\mathbf{O}_{\parallel}\rangle_{\pm}
=
\mp \frac{2\hbar}{5\sqrt{10}}
\frac{\mathcal{F}_{\pm}(k)}
{\mathcal{N}_{\pm}(k)}
\left(
\sin\theta_k,
-\cos\theta_k
\right),
\end{equation}
which exhibits a Rashba type MO texture.
The numerator can alternatively be expanded as
\begin{equation}
\mathcal{F}_{\pm}(k)
=
\sum_{\alpha}
\left[
f_{0,\alpha}
\pm f_{1,\alpha}k
+
\mathcal{O}(k^{2})
\right],
\end{equation}
where
\begin{subequations}
\begin{align}
f_{0,\alpha}
&=
u_{0,\alpha}^{2}
-\frac{3}{\sqrt{2}}
u_{0,\alpha}v_{0,\alpha}
+3v_{0,\alpha}^{2},
\end{align}
\begin{align}
f_{1,\alpha}
&=
\frac{3}{\sqrt{2}}u_{0,\alpha}u_{1,\alpha}
-2u_{0,\alpha}v_{1,\alpha}
-\frac{3}{\sqrt{2}}u_{0,\alpha}w_{1,\alpha} -6u_{1,\alpha}v_{0,\alpha}
+\frac{3}{\sqrt{2}}v_{0,\alpha}v_{1,\alpha}
+v_{0,\alpha}w_{1,\alpha}.
\end{align} 
\end{subequations}

The normalization factor is
\begin{align}
\mathcal{N}_{\pm}(k)
=
\sum_{\alpha}
\Big[
&u_{0,\alpha}^{2}
+
v_{0,\alpha}^{2}
\mp
2k
\left(
u_{0,\alpha}v_{1,\alpha}
+
u_{1,\alpha}v_{0,\alpha}
\right) + \mathcal{O}(k^{2})\Big].
\end{align}

\section{Rashba-Edelstein responses in realistic multiorbital models}

In this section, we present the calculations of the MO Rashba-Edelstein susceptibilities on three multiorbital models of real materials: a typical system that displays the spin Rashba texture (BiTeI), an orbital-Rashba surface alloy (BiAg$_{2}$) and an oxide interface [(111) LaAlO$_{3}$ / SrTiO$_{3}$ bilayer]. These results show that the MO Rashba-Edelstein effect can generically appear.

\begin{figure}[t!]
\includegraphics[width=0.9\textwidth]{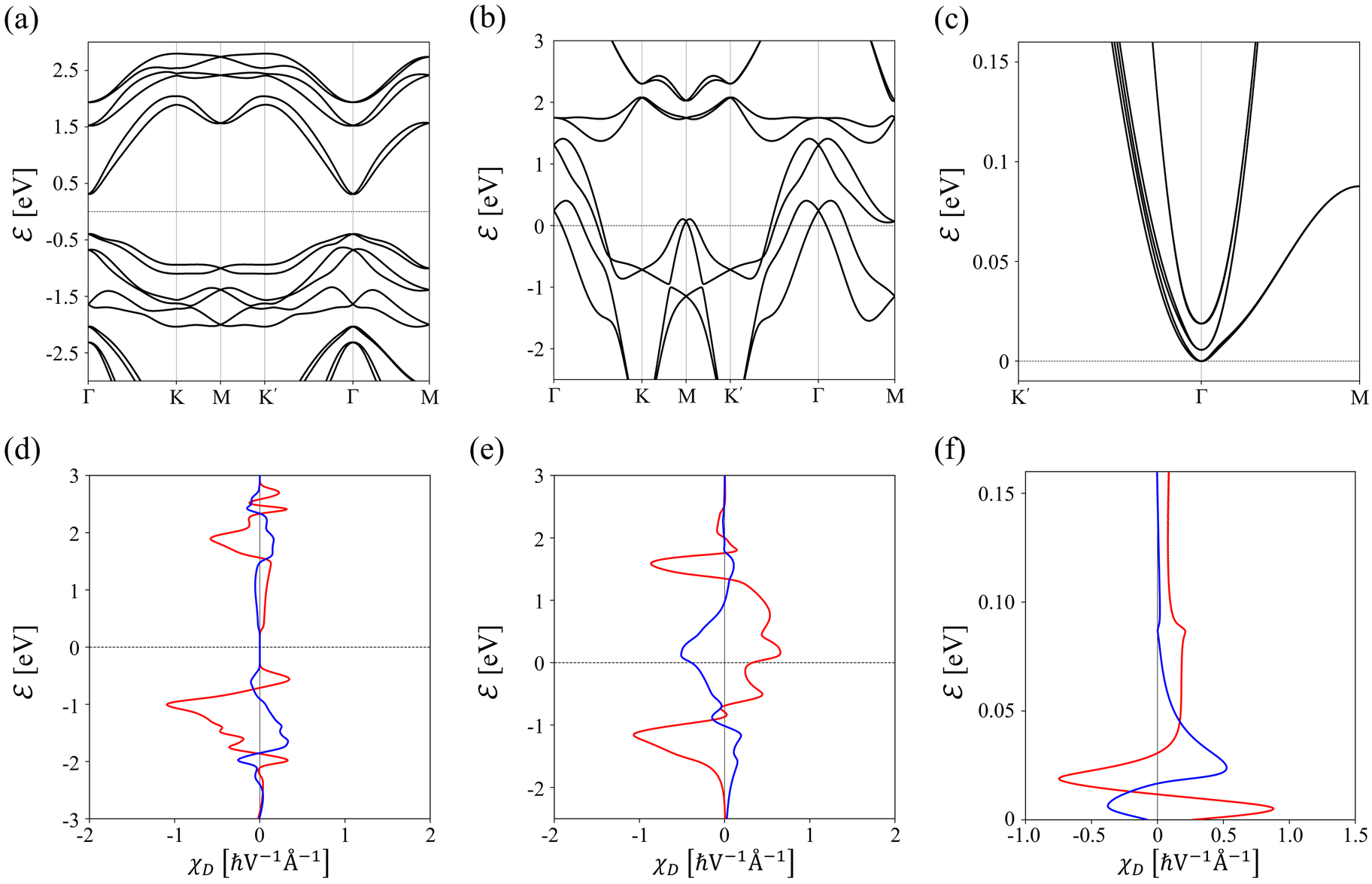}
\caption{\label{fig.s4}
(a)-(c) Band structures of BiTeI, BiAg$_{2}$, and LaAlO$_{3}$/SrTiO$_{3}$ bilayer, respectively. (d)-(f) The spin (red) and MO (blue) Rashba-Edelstein responses for BiTeI, BiAg$_{2}$, and LaAlO$_{3}$/SrTiO$_{3}$ bilayer, respectively.}
\end{figure}

The results are presented in Fig.~\ref{fig.s4}. The upper panels [Fig.~\ref{fig.s4}(a)-\ref{fig.s4}(c)] display the band structures, and the lower panels [Fig.~\ref{fig.s4}(d)-\ref{fig.s4}(f)] show both $\chi_{O}$ and $\chi_{S}$. For each material, the Rashba-Edelstein susceptibilities $\chi_{D}$ ($D = O, S$) are evaluated using the constant-relaxation-time intraband expression in Eq. (10) of the main text. For BiTeI and BiAg$_{2}$, we use $\hbar/\tau = 25$ meV ($\tau = 26.3$ fs). For LaAlO$_{3}$/SrTiO$_{3}$, we use $\tau = 3.4$ ps, following Ref.~\cite{Trama22NM}. The Brillouin-zone integrations were performed using $400 \times 400$ $k$-meshes for BiTeI and BiAg$_{2}$, and a $1000 \times 1000$ $k$-mesh for (111) LaAlO$_{3}$/SrTiO$_{3}$. We use $T = 300$ K for BiTeI and BiAg$_{2}$ and $T = 10$ K for LaAlO$_{3}$/SrTiO$_{3}$. The lattice constant is set to $a = 4.42$ $\text{\AA}$ for BiTeI~\cite{Jahangirzadeh21PRm} and $a = 5.011$ $\text{\AA}$ for BiAg$_{2}$~\cite{Go17SR}. For LaAlO$_{3}$/SrTiO$_{3}$, we use the cubic SrTiO$_{3}$ lattice constant $a_{0} = 3.905$ $\text{\AA}$, corresponding to the triangular in-plane lattice constant is set to $a = \sqrt{2}a_{0} = 5.522$ $\text{\AA}$~\cite{Trama22NM}.

First, we consider BiTeI, a representative strong-SOC Rashba system. To describe BiTeI, we employ the corresponding 24-band $sp$ tight-binding model introduced in Ref.~\cite{Ganjehie24PRb}. The result [Fig.~\ref{fig.s4}(d)] shows that this conventional spin-Rashba texture is naturally accompanied by a finite MO response. This implies that the MO Rashba effect is not restricted to specially designed models, but arises naturally in an ordinary spin-Rashba material.

Next, we consider the results of BiAg$_{2}$, in which the orbital degree of freedom plays a central role. To describe BiAg$_{2}$, we employ the 10-band $sp$ tight-binding model introduced in Ref.~\cite{Go17SR}. The result in Fig.~\ref{fig.s4}(e) reveals a MO-dominant regime for $-0.91 \leq \mathcal{E} \leq -0.80$ eV, where $\chi_{O}$ remains sizable while $\chi_{S}$ is strongly suppressed. This provides a realistic counterpart of the behavior found in the toy models introduced in the main text, showing that the MO channel can remain sizable while the spin channel is suppressed. A second and broader MO-dominant regime is found for $-0.22 \leq \mathcal{E} \leq 0.04$ eV, which includes the Fermi energy. Throughout this interval, $\chi_{O}$ remains sizable and exceeds $\chi_{S}$.

Finally, we consider the results of (111) LaAlO$_{3}$/SrTiO$_{3}$ bilayer. We employ a 12-band $t_{2g}$ tight-binding model to describe the (111) LaAlO$_{3}$/SrTiO$_{3}$ interface introduced in Ref.~\cite{Trama22NM}. In this model, $\chi_{S}$ undergoes a strong filling-dependent suppression due to multiband competition, whereas $\chi_{O}$ remains finite in the same energy range ($23 \leq \mathcal{E} \leq 53$ meV) [Fig.~\ref{fig.s4}(f)].

The model calculations for the three real materials support the robustness of the MO Rashba-Edelstein effect across qualitatively different microscopic platforms. BiTeI shows that an MO Rashba-Edelstein effect naturally accompanies the conventional spin Rashba-Edelstein effect in a strong-SOC Rashba system. BiAg$_{2}$ and LaAlO$_{3}$/SrTiO$_{3}$ demonstrate that the MO Rashba-Edelstein effect can remain sizable even though the spin channel is selectively suppressed.


\end{document}